\documentclass[twocolumn,trackchanges,twocolappendix]{aastex631}

\usepackage{natbib}
\usepackage{epsfig}
\usepackage{graphicx}
\usepackage{subfigure}
\usepackage{float}
\usepackage{amsmath}
\usepackage{color}
\usepackage{amssymb}
\usepackage{apjfonts}
\usepackage{multirow}
\usepackage{colortbl}
\usepackage{ulem}

\newcommand{\PMO}{Purple Mountain Observatory, Chinese Academy of Sciences, Nanjing 210023, China}
\newcommand{\USTC}{School of Astronomy and Space Sciences, University of Science and Technology of China, Hefei 230026, China}

\shortauthors{Wang et al.}

\graphicspath{{./}{figures/}}

\usepackage{amsmath}
\begin{document}

\title{Measuring the Angular Auto-power Spectrum of Fast Radio Burst Dispersion Measures as a Cosmological Probe and Baryon Tracer}


\correspondingauthor{Jun-Jie Wei, Xue-Feng Wu}
\email{jjwei@pmo.ac.cn, xfwu@pmo.ac.cn}

\author[0000-0003-3635-5375]{Bao Wang}
\affiliation{\PMO}
\affiliation{\USTC}

\author[0009-0001-3701-6650]{Zhiyu Lu}
\affiliation{\USTC}
\affiliation{CAS Key Laboratory for Research in Galaxies and Cosmology, School of Astronomy and Space Science, University of Science and Technology of China, Hefei, Anhui 230026, China}
	
\author[0000-0003-2721-2559]{Yang Liu}
\affiliation{\PMO}

\author[0000-0003-0162-2488]{Jun-Jie Wei}
\affiliation{\PMO}
\affiliation{\USTC}

\author[0000-0002-6299-1263]{Xue-Feng Wu}
\affiliation{\PMO}
\affiliation{\USTC}

\begin{abstract}
Fluctuations in the cosmic electron density are imprinted on the dispersion measures (DMs) of fast radio bursts (FRBs), making DMs a promising probe of cosmology and the spatial distribution of ionized baryons.
In this work, we present the first measurement of the angular auto-power spectrum of FRB DMs, using 3455 apparently non-repeating bursts from the CHIME/FRB Catalog 2. We detect an angular correlation signal at $\sim 2.8 \sigma$ significance, associated with large-scale electron-density fluctuations. 
By fitting the measured spectrum to theoretical models, we constrain two key parameter combinations: $\Omega_{\rm b}h^2$--$H_0$, which probes the cosmic baryon density and expansion rate, and $\Omega_{\rm b}h^2$--$f_{\rm d}$, which traces the baryon fraction in cosmic large-scale structure (LSS). 
We further assess the sensitivity of the power-spectrum method to systematic uncertainties arising from the assumed FRB redshift distribution and from the DM contributions of host galaxies (${\rm DM}_{\rm host}$), the Galactic halo (${\rm DM}^{\rm MW}_{\rm halo}$), and the Milky Way interstellar medium (${\rm DM}^{\rm MW}_{\rm ISM}$), using mock samples. 
Our results demonstrate that the angular power spectrum is largely insensitive to uncorrelated DM components such as ${\rm DM}_{\rm host}$, thereby effectively mitigating the impact of poorly constrained host-galaxy systematics.
In contrast to the mean ${\rm DM}_{\rm LSS}$--$z$ relation, this method does not require individual redshift measurements but relies only on the overall redshift distribution, and partially breaks the parameter degeneracies in the $\Omega_{\rm b}h^2$--$H_0$ and $\Omega_{\rm b}h^2$--$f_{\rm d}$ planes.
These findings establish the DM angular power spectrum as a reliable cosmological probe and a powerful baryon tracer.
\end{abstract}


\section{Introduction}
\label{Sec:Intro}

Mapping the spatial distribution of ionized baryons is a key observational challenge in understanding the formation and evolution of cosmic structures. 
Although the total cosmic baryon budget has been precisely constrained by measurements of the cosmic microwave background (CMB) \citep{Planck2020A&A}, the late-time spatial distribution of baryons is still highly uncertain \citep{Fukugita1998ApJ, Shull2012ApJ, McQuinn2016ARA&A}. 
The majority of baryons are expected to reside in diffuse ionized gas associated with the intergalactic medium (IGM), the circumgalactic medium (CGM), and the cosmic web, yet these components remain inherently difficult to probe directly. 
Existing observational constraints, derived from X-ray observations \citep{Nicastro2018Natur, Zhang2024A&A}, the thermal Sunyaev--Zel'dovich (SZ) effect \citep{Tanimura2019MNRAS, Graaff2019A&A}, and the kinetic SZ effect \citep{Schneider2022MNRAS, RiedGuachalla2025PhRvD, Hadzhiyska2025PhRvD, Sunseri2026MNRAS}, have provided important insights into diffuse ionized gas. However, these tracers are predominantly sensitive to dense plasma, particularly in galaxy clusters, leaving the more diffuse structures poorly constrained. 

Fast radio bursts (FRBs) are bright, millisecond-duration radio transients \citep{Lorimer2007Sci, Xiao2021SCPMA, Petroff2022A&ARv, Zhang2023RvMP}. 
Their dispersion measures (DMs), which quantify the integrated electron column density along the line of sight, establish FRBs as powerful cosmological probes and sensitive tracers of the ionized baryon content \citep{Bhandari2021Univ, Wu2024ChPhL, Wang2026SSPMA}.
In particular, the redshift evolution of the cosmic large-scale structure (LSS) contribution to the DM (including the IGM and the circumgalactic medium of intervening halos), denoted as ${\rm DM}_{\rm LSS}$, is encapsulated in the well-established ${\rm DM}_{\rm LSS}$--$z$ relation \citep{Ioka2003ApJ, Deng2014, Macquart2020Natur}. This relation has been widely exploited in a variety of cosmological and baryon-census studies, providing critical constraints on the Hubble constant \citep{Hagstotz2022MNRAS, James2022MNRAS, Wu2022MNRAS, Zhao2022arXiv, Liu2023ApJ, 2023ApJ...955..101W, JahnsSchindler2025PhRvD, Xu2025ApJ, Gao2025A&A, Sales2025arXiv, Zhuge2026ApJ, Liu2026ApJb}, dark energy \citep{Zhou2014PhRvD, Walters2018ApJ, 2018ApJ...860L...7W, Zhao2020ApJ, Wang2025ApJ, Yan2025ChPhC, Ribeiro2026arXiv}, and the cosmic (or IGM) baryon abundance \citep{Li2019ApJ, Li2020MNRAS, Wei2019JCAP, Macquart2020Natur, Yang2022ApJ, Wang2023ApJ, Zhang2025ApJ, Lemos2025JCAP, Zhang2025arXiv, Sales2026A&A}. 

However, the cosmological application of FRBs is currently hampered by several key factors: 
(i) the limited number of FRBs with precise redshift measurements; 
(ii) the inherent degeneracy between cosmological and baryonic parameters in the ${\rm DM}_{\rm LSS}$--$z$ relation; 
and (iii) systematic uncertainties associated with estimating the various components of the observed DM.  
Accurately disentangling the DM contributions from the host galaxy, the Milky Way halo, and the Galactic interstellar medium (ISM) remains a substantial challenge. 
A common approach is to model these components using probability distributions \citep{Macquart2020Natur, Zhang2020ApJ, Zhang2021ApJ, Takahashi2021MNRAS}. 
However, this method cannot fully capture the sightline-to-sightline diversity and local environmental variations, which can still introduce biases in parameter inference \citep{Wang2025ApJ, Xu2025ApJ, Lemos2026arXiv}.
These uncertainties therefore represent a common obstacle to precise cosmology with FRB DMs. 

The cosmic DM provides a line-of-sight integrated tracer of the cosmic free-electron density field. 
Fluctuations in the cosmic electron density are imprinted on the projected FRB DM field, and the angular auto-power spectrum of the DM field provides a two-point statistical description of these fluctuations
\citep{Masui2015PhRvL, Shirasaki2017PhRvD}.
This power-spectrum method offers several anticipated advantages: (i) it does not require precise redshifts for individual FRBs, although redshift information remains important; (ii) the scale-dependent information helps break degeneracies among parameters; and (iii) it is relatively insensitive to scale-independent DM components, such as those arising from local environments, which primarily increase the noise and statistical uncertainties rather than the angular correlations. 
Therefore, the DM power spectrum has the potential to serve as a reliable cosmological probe. 
Several studies have presented forecasts for its application, including probes of baryonic feedback \citep{Wayland2026MNRAS, Sharma2026ApJa, Reischke2026JCAP, Feng2026ApJ}, cosmic reionization \citep{Dai2021JCAP}, the cosmic growth rate \citep{Wang2025EPJC} and tests of gravity \citep{Neumann2025OJAp, Zhou2026PhRvD}. 
More recently, cross-correlation measurements of the DM field with galaxy distributions, the SZ effect, and X-ray emission have provided observational evidence that the projected DM field traces the large-scale structure \citep{Wang2026PRL,Takahashi2026PASJ, Reischke2026OJAp, Sharma2026NatAs, Sharma2026ApJc}. 
Nevertheless, a direct measurement of the DM auto-power spectrum from real FRB data remains absent. 
Such a measurement is challenged not only by the limited size of FRB samples, but also by systematic effects that are difficult to quantify and that require careful treatment for a reliable cosmological interpretation. 
These include uncertainties in the modeling of the electron density power spectrum, such as those arising from baryonic feedback \citep{Sharma2026ApJa, Wang2026PRL} and the electron bias \citep{Andrew2026arXiv}.

In this work, we present the first measurement of the angular auto-power spectrum of FRB DMs, using 3455 non-repeating bursts from the CHIME/FRB Catalog 2.
After subtracting the Milky Way ISM foreground, we construct a residual DM field and measure its angular power spectrum.
We detect an auto-correlation signal at a significance level of $\sim 2.8\sigma$.
We then fit the measured bandpowers to the theoretical predictions, constraining two parameter combinations: $\Omega_{\rm b}h^2$--$H_0$ as a cosmological probe, and $\Omega_{\rm b}h^2$--$f_{\rm d}$ as a baryon tracer.
Furthermore, 
we test the sensitivity of our constraints to variations in the assumed FRB redshift distribution, host-galaxy DM contributions, and the modeling of the Milky Way halo and ISM, thereby establishing the reliability of this approach when these systematic uncertainties are properly accounted for. 

This paper is structured as follows.
In Section~\ref{Sec:Dispersion}, we describe the theoretical framework for the FRB DM angular power spectrum. 
In Section~\ref{Power measured}, we present the CHIME/FRB sample and the measured bandpowers.
In Section~\ref{parameters}, we derive the cosmological and baryonic constraints from the measured values.
In Section~\ref{test}, we examine the sensitivity of our results to various systematic uncertainties.
Finally, we summarize our conclusions and discuss future prospects in Section~\ref{Sec:Conclusion}.
Throughout this paper, we adopt a flat $\Lambda$CDM cosmology as the fiducial model, with $H_0 = 67.74 \, \mathrm{km\; s^{-1}\; Mpc^{-1}}$, $\Omega_{m} = 0.315$, and $\Omega_{b}h^2= 0.0224$ \citep{Planck2020A&A}.


\section{DM Angular Power Spectrum}
\label{Sec:Dispersion}

The observed DM of an FRB is defined as the line-of-sight integral of the free-electron density, ${\rm DM} \equiv \int n_e \, {\rm d} l$. 
For a burst at redshift $z$, the observed DM can be decomposed as
\begin{equation}\label{eq:dm_decomp}
	{\rm DM}_{\rm obs} = {\rm DM}^{\rm MW}_{\rm ISM} + {\rm DM}^{\rm MW}_{\rm halo}
	+ {\rm DM}_{\rm LSS} (z)+ \frac{{\rm DM}_{\rm host}}{1+z},
\end{equation}
where the terms on the right-hand side denote the contributions from the Milky Way ISM (${\rm DM}^{\rm MW}_{\rm ISM}$), the Milky Way halo (${\rm DM}^{\rm MW}_{\rm halo}$), the cosmic LSS (${\rm DM}_{\rm LSS}$), and the host galaxy including the local environment (${\rm DM}_{\rm host}$), respectively. 
The factor $1/(1+z)$ converts the rest-frame host-galaxy DM to the observed frame. 

In this work, we focus on the fluctuations of the intergalactic DM field as projected across different lines of sight on the sky. 
The projected field can be separated into a redshift-dependent mean component and a fluctuation component sourced by the cosmic LSS. 
The angular power spectrum of these DM fluctuations provides a statistical measure of their variance as a function of angular scale. 
For a source at redshift $z$, the cosmic LSS contribution can be written as
\begin{equation}\label{eq:dm_igm}
	\begin{split}
	{\rm DM}_{\rm LSS}(\hat{\boldsymbol n},z)
	& = \int_0^{z} dz^{\prime}\, W_{\rm LSS}(z^{\prime}) \left[1+\delta_e(\hat{\boldsymbol n},z^{\prime})\right] \\
	& = \left\langle\mathrm{DM}_{\mathrm{LSS}}\right\rangle(z) 
	+ \delta {\rm DM_{\mathrm{LSS}}}(\hat{\boldsymbol n}, z) ,
	\end{split}
\end{equation}
where $\delta_e$ is the dimensionless fluctuation of the free-electron density.
The first term on the right-hand side of the second line corresponds to the average LSS contribution, which defines the mean ${\rm DM}_{\rm LSS}$--$z$ relation \citep{Ioka2003ApJ, Deng2014},
while the second term represents the DM fluctuation induced by LSS along the line of sight.
The weighting function for ${\rm DM}_{\rm LSS}$ is given by
\begin{equation}\label{eq:dmlss_weight}
	W_{\rm LSS}(z) = \frac{c\,\bar n_{b,0}}{H(z)} \, f_{\rm d}(z) \chi_e(z) (1+z).
\end{equation}
Here $\bar n_{b,0}=3H_0^2\Omega_b/(8\pi G m_p)$ is the present-day mean baryon number density, with $m_p$ the proton mass, $H_0$ the Hubble constant, and $\Omega_{b}$ the present-day baryon density parameter. 
For a flat $\Lambda$CDM cosmology, the Hubble expansion rate is 
$H(z)=H_0\sqrt{\Omega_m(1+z)^3+1-\Omega_m}$, where $\Omega_{m}$ denotes the present-day matter density parameter. 
The factor $f_{\rm d}(z)$ represents the fraction of diffuse ionized baryons in the cosmic LSS, and $\chi_e(z)=\frac{3}{4} \chi_{e,\mathrm{H}}(z)+\frac{1}{8}\chi_{e, \mathrm{He}}(z)$ is the number of free electrons per baryon, determined by the ionization fractions of hydrogen ($\chi_{e,\mathrm{H}}(z)$) and helium ($\chi_{e,\mathrm{He}}(z)$). 
At redshift $z<3$, both hydrogen and helium are fully ionized \citep{Meiksin2009RvMP, Becker2011}, implying $\chi_{e}=7/8$. 
As massive halos are more abundant in the late Universe, $f_{\rm d}$ is expected to evolve with redshift \citep{Meiksin2009RvMP}. 
However, the redshift dependence of $f_{\rm d}$ remains poorly constrained by current FRB data \citep{Wang2023ApJ, Liu2026ApJb}; we therefore adopt a constant $f_{\rm d}$ as an approximation in our analysis.

The projected ${\rm DM}_{\rm LSS}$ fluctuation field is obtained by integrating the free-electron density fluctuations along the lines of sight toward the FRB population, and can be written as
\begin{equation}\label{eq:projected_dm}
	\delta {\rm DM}_{\rm LSS}^{\rm proj}(\hat{\boldsymbol n})
	= \int dz\, {\cal W}_{\rm LSS }(z)\, \delta_e(\hat{\boldsymbol n},z),
\end{equation}
where the projection weight function is
\begin{equation}\label{eq:dm_weight}
	{\cal W}_{\rm LSS }(z) = W_{\rm LSS}(z) \int_z^\infty dz^{\prime}\,p_s(z^{\prime}).
\end{equation}
Here $p_s(z)$ denotes the normalized redshift distribution of the FRB sample.
In principle, this distribution could be empirically reconstructed from secure host-galaxy identifications. However, given the current paucity of FRBs with reliable redshift measurements, combined with observational uncertainties and selection biases, a direct empirical determination remains highly challenging. We therefore adopt a phenomenological parameterization,
$p_s(z) \propto z^2 \exp(-\alpha z)$ with $\alpha = 3.5$,
as motivated by Refs.~\citep{Rafiei2020PhRvD, Sharma2026ApJa}.
This form effectively captures the competition between the increasing comoving volume at low redshift and the decreasing detectability of FRBs at higher redshift, which arises from a combination of instrumental selection effects, flux limits, and possible intrinsic evolution of the FRB population. Within this framework, redshift information enters the angular power spectrum through the assumed form of $p_s(z)$ in the projection weight, without requiring individual FRB redshift measurements. This allows us to fully exploit large samples of unlocalized FRBs. 
Nevertheless, redshift information remains important, and the lack of FRB redshifts remains one of the principal challenges for cosmological applications of FRBs.
Individual redshift measurements would enable tomographic analyses that separate contributions from different cosmic epochs, whereas their absence leads to a loss of radial information, larger statistical uncertainties, and increased sensitivity to redshift-distribution systematics.
We assess the sensitivity of our results to variations in the assumed $p_s(z)$ in Section~\ref{test}.

Expanding the projected field in spherical harmonics,
\begin{equation}
	\delta{\rm DM}_{\rm LSS}^{\rm proj}(\hat{\boldsymbol n}) 
	= \sum_{\ell m} a_{\ell m}  Y_{\ell m}(\hat{\boldsymbol n}), 
\end{equation}
the ${\rm DM}_{\rm LSS}$ auto-correlation angular power spectrum is defined as
\begin{equation}\label{eq:cl}
	C_\ell^{\rm LSS-LSS} = \frac{1}{2\ell+1}
	\sum_{m=-\ell}^{\ell} \left|a_{\ell m} \right|^2 .
\end{equation}
The procedure for extracting $\delta{\rm DM}^{\rm proj}_{\rm LSS}(\hat{\boldsymbol n})$ from Eq.~\eqref{eq:dm_decomp} is detailed in Section~\ref{Power measured}.
Under the Limber approximation, the power spectrum is given by \citep{Shirasaki2017PhRvD, Dai2021MNRAS, Sharma2026ApJa}
\begin{equation}\label{eq:cl_igm}
	C_\ell^{\rm LSS-LSS}
	= \int {\rm d} z\, \frac{H(z)}{c\,\chi^2(z)}
	{\cal W}_{\rm LSS}^2(z)
	P_{ee} \left( k=\frac{\ell+1/2}{\chi(z)},z \right),
\end{equation}
where $\chi(z)$ is the comoving distance and $P_{ee}(k,z)$ is the three-dimensional power spectrum of the electron density fluctuation $\delta_e$. 
Assuming a linear bias relation, the electron power spectrum can be expressed as $P_{ee}(k,z)=b_e^2P_m(k,z)$, where $P_m(k,z)$ is the matter power spectrum and $b_e$ denotes the electron bias factor. 
On large scales, the electron density field is expected to trace the underlying matter distribution, corresponding to $b_e\simeq 1$ \citep{Takahashi2021MNRAS}.
We therefore adopt $b_e = 1$ throughout our analysis, while noting that the nonlinear matter power spectrum is computed using  the Mead2020 feedback model \citep{Mead2021MNRAS} as implemented in \textsc{camb} \citep{Lewis2000ApJ}, with the baryonic feedback parameter fixed to $\log_{10} (T_{\rm AGN}/{\rm K})=7.8$.
The impact of $T_{\rm AGN}$ on the DM power spectrum is discussed in Appendix~\ref{app:baryon}.

The extragalactic DM fluctuation field also contains contributions from host galaxies.
The extragalactic DM power spectrum can thus be decomposed as \citep{Shirasaki2017PhRvD}
\begin{equation}\label{eq:cl_DM}
	\begin{split}
	C_\ell^{\rm DM} & = C_\ell^{\rm LSS-LSS} +
	C_\ell^{\rm host-host} + C_\ell^{\rm LSS-host} \\
	& \simeq C_\ell^{\rm LSS-LSS}. 
	\end{split}
\end{equation}
The amplitudes of host–host and LSS–host terms depend on the normalized source redshift distribution and the mean ${\rm DM_{host}}$.
This approximation is justified by previous studies showing that $C_\ell^{\rm host-host}$ is more than two orders of magnitude smaller than $C_\ell^{\rm LSS-LSS}$, while $C_\ell^{\rm LSS-host}$ is smaller by more than one order of magnitude \citep{Shirasaki2017PhRvD, Dai2021MNRAS}.
Given that these host-related contributions are subdominant and not distinguishable at the current precision level, we neglect them in the signal modeling. 

Nevertheless, stochastic contributions from host galaxies, their local environments, the Milky Way halo, and residual foreground subtraction errors do not correlate across different sightlines; they therefore contribute only to the measurement variance, manifesting as an approximately scale-independent white-noise term \citep{Shirasaki2017PhRvD}. 
The spectrum can accordingly be written as
\begin{equation}\label{eq:cl_noise}
	\hat C_\ell^{\rm DM} = C_\ell^{\rm DM} + N_\ell^{\rm DM},
\end{equation}
where the shot-noise power spectrum is given by
\begin{equation}\label{eq:shot_noise}
	N_\ell^{\rm DM}	\simeq \frac{\sigma_{\rm DM}^2}{\bar n_{\rm FRB}} .
\end{equation}
Here $\bar n_{\rm FRB}$ is the sky-projected number density of FRBs, and
$\sigma_{\rm DM}$ includes the combined scatter arising from host galaxies, source
environments, Milky Way halo contributions, and residual systematic uncertainties in foreground subtraction. 
Since these stochastic residuals are assumed to be statistically independent among different FRB sightlines, they do not generate angular correlations and can be fully characterized by this single white-noise parameter $\sigma_{\rm DM}$.
Importantly, this treatment does not require detailed probability distributions for each individual DM component. This inherent property renders the DM angular power spectrum substantially less sensitive to scale-independent DM systematics than the mean ${\rm DM}_{\rm LSS}$--$z$ relation approach.

FRB localization uncertainties can suppress the measured DM power spectrum on small angular scales \citep{RafieiRavandi2021ApJ, Wang2026PRL, Zhai2026arXiv}. 
We model this effect with a Gaussian
window function as $C_{\ell}^{\rm DM,loc}
=C_{\ell}^{\rm DM}\exp{[-\ell^2/(2\ell_{\rm loc}^2)]}$, 
where $\ell_{\rm loc}$ characterizes the angular scale of the FRB localization uncertainty.
For the first CHIME/FRB catalog, this scale is estimated to lie in the range $315\lesssim \ell_{\rm loc}\lesssim1396$ \citep{RafieiRavandi2021ApJ}.


\section{Power Spectrum Measurement}\label{Power measured}

\subsection{DM Map}

\begin{figure}
	\includegraphics[keepaspectratio,clip,width=0.5\textwidth]{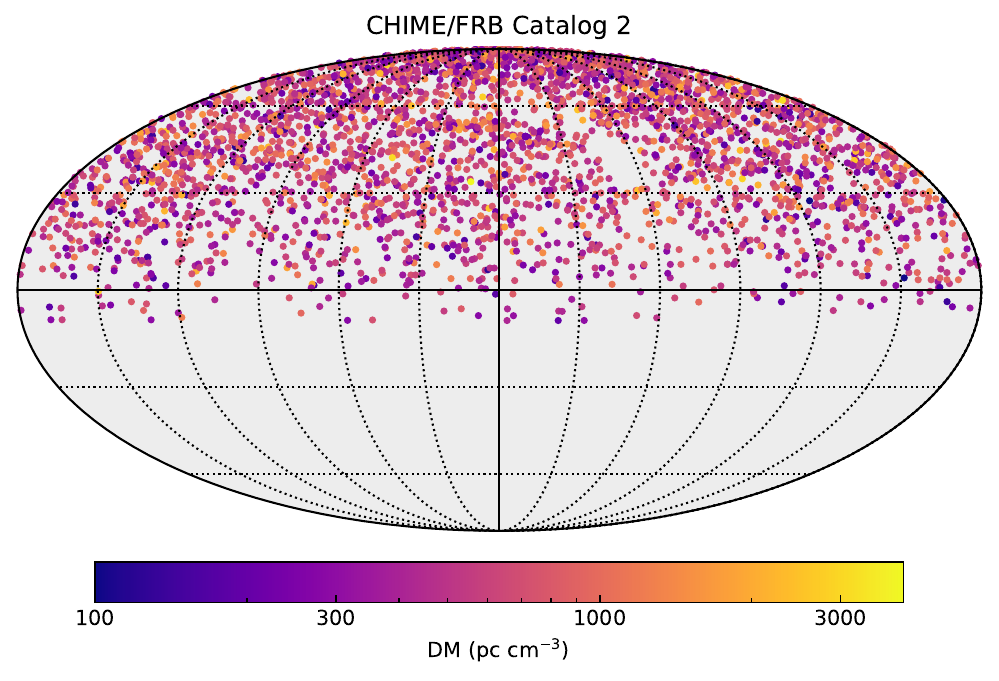}
	\caption{Sky distribution of the 3455 apparently non-repeating FRBs from the second CHIME/FRB catalog. The points are plotted in equatorial coordinates (Mollweide projection) and color-coded by their observed DMs.}
	\label{Fig1}
\end{figure}

The power spectrum measurement presented in this work is based on the second CHIME/FRB catalog \citep{Chime2026ApJSa}, which reports sky positions and observed DMs for 4539 FRBs. 
We exclude repeaters and events flagged for exclusion from parameter inference due to non-nominal telescope operation. 
The resulting sample comprises 3455 apparently non-repeating FRBs, which are primarily concentrated in the northern sky at declinations $\delta > -10^{\circ}$. 
The mean number density $\bar{n}_{\rm FRB}$ of the sample in this region is approximately $0.17~\;{\rm deg}^{-2}$.
Figure~\ref{Fig1} shows the sky distribution of the selected sample, with the color of each point indicating its observed DM. 

To construct the DM fluctuation map, we first perform a Galactic foreground correction by subtracting the Milky Way ISM contribution from each burst. 
The values of ${\rm DM}^{\rm MW}_{\rm ISM}$ are estimated using the NE2025 electron density model \citep{Ocker2026ApJ}. 
For the $i$-th FRB, we define the residual DM by subtracting both the Milky Way ISM contribution and the mean of the foreground-corrected DM field:
\begin{equation}\label{eq:residual_dm}
	\Delta {\rm DM}_i = {\rm DM}_{{\rm obs},i} - {\rm DM}^{\rm MW}_{{\rm ISM},i}
	- \left\langle {\rm DM}_{\rm obs}-{\rm DM}^{\rm MW}_{\rm ISM} \right\rangle ,
\end{equation}
where the angle brackets denote the average over the full sample. 
This mean subtraction effectively removes the monopole component, so that $\Delta{\rm DM}$ primarily traces spatial variations in the foreground-corrected DM field.
We further assume that ${\rm DM}^{\rm MW}_{\rm halo}$ is approximately isotropic across the sky; consequently, its mean is largely absorbed by this subtraction step and does not bias the fluctuation measurement. 
The possible impact of the anisotropic Milky Way halo is addressed in Section~\ref{test}.

\subsection{Power Spectrum}

We estimate the DM angular power spectrum using the catalog-based estimator implemented in \textsc{NaMaster}\footnote{\url{https://namaster.readthedocs.io/en/latest/4Catalogs.html}} \citep{Alonso2019MNRAS}. 
Specifically, we create an \texttt{NmtFieldCatalog} \citep{Wolz2025JCAP} object from the angular positions of the FRBs and their residual DM values, $\Delta {\rm DM}_i$ as defined in Equation~(\ref{eq:residual_dm}), assigning equal weights to all bursts. 
This catalog-based estimator operates directly on the ungridded FRB positions, thereby circumventing the need to choose an arbitrary HEALPix resolution \citep{Gorski2005ApJ} that could introduce spurious pixelisation effects in the sparse DM field.
The auto-power spectrum is computed from the catalog field using the \texttt{compute\_coupled\_cell}. 
However, the incomplete and anisotropic sky coverage of the CHIME/FRB survey couples different multipoles, such that the measured pseudo-spectrum can be written as $\tilde{C}_{\ell}=\sum_{\ell^{\prime}} M_{\ell \ell^{\prime}} C_{\ell^{\prime}}$, where $C_{\ell^{\prime}}$ is the underlying true spectrum and $M_{\ell \ell^{\prime}}$ is the mode-coupling matrix determined by the survey window \citep{Wolz2025JCAP, Wolz2026arXiv, Cheng2026ApJ}.
The matrix $M_{\ell \ell^{\prime}}$ quantifies how power at a true multipole $\ell^{\prime}$ is mixed into the observed multipole $\ell$ by the survey window. 
For the catalog-based estimator, this window is determined from the discrete FRB positions and their weights. 
We therefore construct an \texttt{NmtWorkspace} object to deconvolve the window effect from the measured spectrum.

\begin{figure}
	\includegraphics[width=0.47\textwidth]{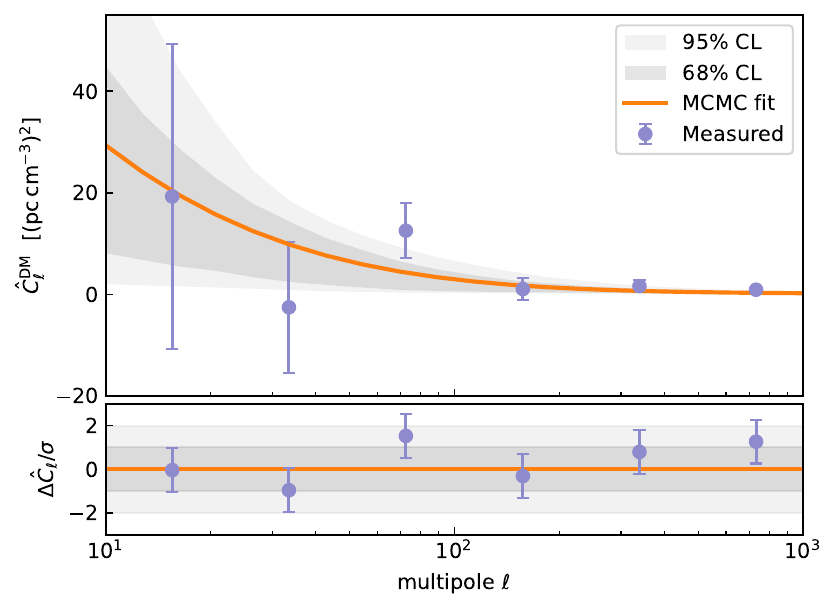}
	\caption{
        Measured DM angular power spectrum from the second CHIME/FRB Catalog. 
		The purple points show the bandpowers with $1\sigma$ Gaussian uncertainties. 
		The orange curve represents the best-fit spectrum from the MCMC analysis (see Section~\ref{parameters}); the dark and light gray shaded regions denote the 68\% and 95\% confidence intervals derived from the MCMC posterior distributions, respectively. 
		The lower panel displays the normalized residuals, $\Delta \hat{C}_\ell/\sigma$, relative to the best-fit model. 
	}
\label{Fig2}
\end{figure}

The final bandpowers are estimated over the multipole range $10 \leq \ell \leq 1000$, divided into six logarithmically spaced bins. 
We also repeat the analysis using different numbers of multipole bins, and find that the signal remains stable under various cases (see the details in Appendix~\ref{app:binning}).
The lower cutoff excludes large angular scales ($\ell < 10$), where the measurement is susceptible to systematic effects arising from the survey window and non-uniform coverage.
A null test presented in Appendix~\ref{app:null} shows larger scatter and larger deviations from zero at low multipoles.
The resulting DM angular power spectrum is presented in Figure~\ref{Fig2}, with uncertainties estimated via the analytical Gaussian covariance \citep{Wolz2026arXiv}.
To evaluate the statistical significance of the measured angular correlations, we perform a $\chi^2$-test following \cite{Wang2026PRL}.
We find a detection significance of $\sim 2.8\sigma$ relative to the null spectrum, as detailed in Appendix~\ref{app:significance}.

\begin{figure*}
	\centering
	\includegraphics[width=0.48\textwidth]{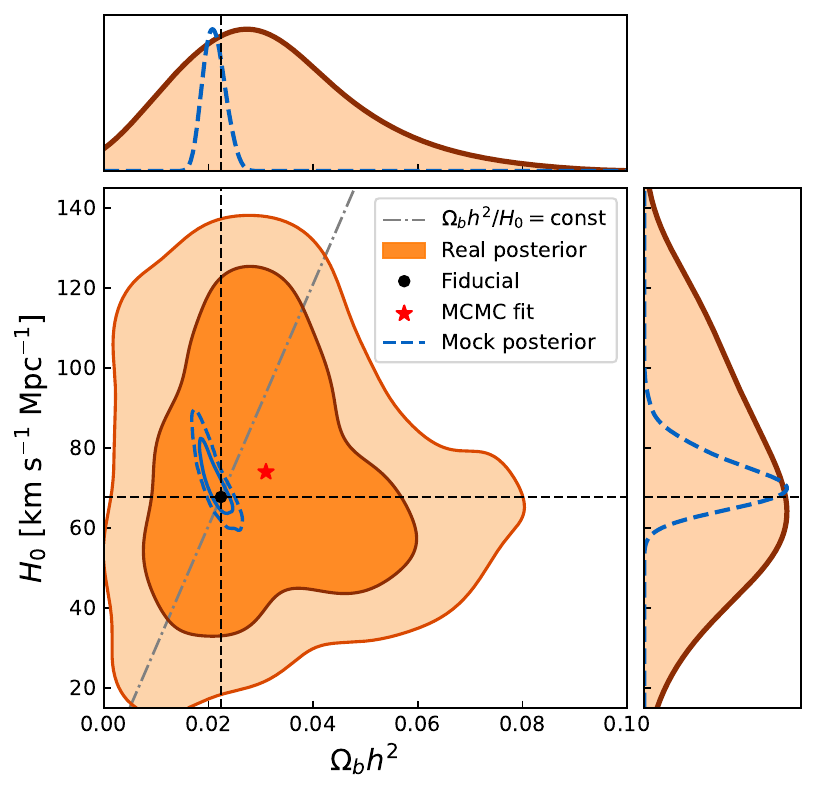}
	\includegraphics[width=0.48\textwidth]{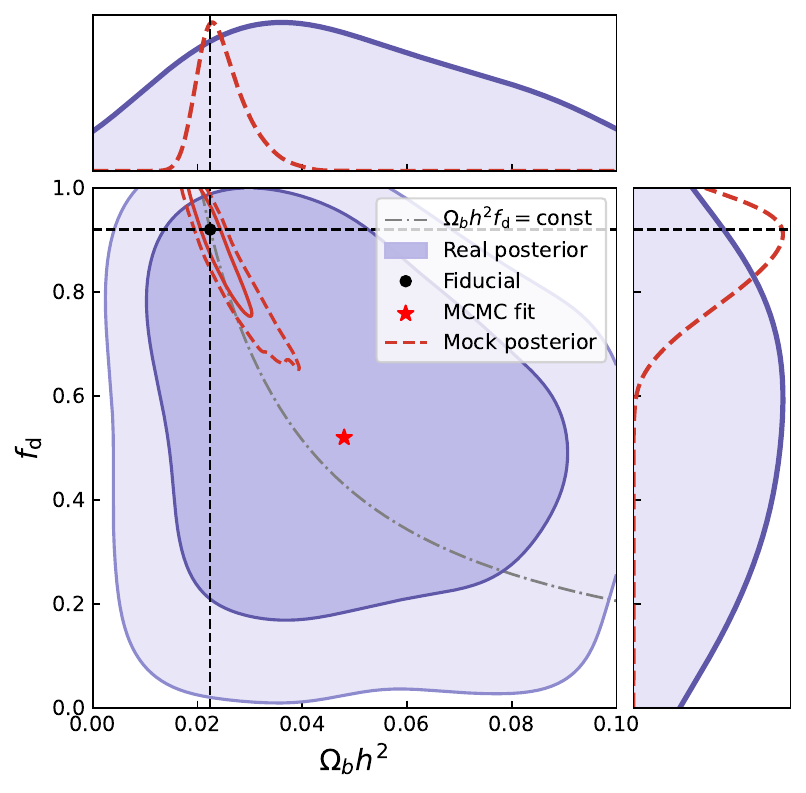}
	\vskip-0.1in
	\caption{Posterior constraints derived from the measured DM angular power spectrum.
		The left and right panels show the constraints in the $\Omega_{\mathrm{b}}h^2$--$H_0$ and $\Omega_{\mathrm{b}}h^2$--$f_{\rm d}$ parameter spaces, respectively.
		The filled contours show the posterior distributions obtained from the real CHIME/FRB sample, while the dashed contours show the constraints derived from $10^5$ mock FRB samples generated from the fiducial model described in Section~\ref{test}.
		In both panels, the black dots and black dashed lines indicate the fiducial input values, and the red stars mark the best-fit values from the MCMC analysis.
		The gray dot-dashed lines illustrate the degeneracy directions expected from the mean ${\rm DM}_{\rm LSS}$--$z$ relation, corresponding to $\Omega_{\mathrm{b}}h^2/H_0={\rm const}$ and $\Omega_{\mathrm{b}}h^2 f_{\rm d}={\rm const}$, respectively.}
	\label{Fig3}
\end{figure*}

~\\
\subsection{Selection Effects}

Selection effects in CHIME/FRB observations have been extensively studied \citep{James2022MNRAS, Merryfield2023AJ, Cheng2026ApJ, Ribeiro2026arXivb}, but their impact on the DM angular auto-power spectrum remains poorly understood.
In this work, we investigate several potential effects, including Galactic foreground effects, declination-dependent sensitivity, and high-DM detection bias (see Appendix~\ref{app:selection} for details). 
The declination-dependent sensitivity of CHIME/FRB may lead to nonuniform FRB detections across the sky \citep{Chime2026ApJSa}.
Previous simulations have also shown that selection against high-DM FRBs can suppress the DM--galaxy cross-power spectrum \citep{Cheng2026ApJ}, which motivates our investigation of its potential impact on the DM auto-power spectrum.
We find that stricter Galactic latitude cuts generally suppress the measured bandpowers across most angular scales, although the differences remain within the statistical uncertainties.
Since these variations do not provide compelling evidence for significant Galactic contamination, we retain the full sample to maximize the statistical power of our measurement.
Meanwhile, declination cuts have little impact on the measured spectrum, except for minor variations at the low multipoles.
Excluding the highest 10\% and 20\% of observed DM sources leads to noticeable changes in several bandpowers, particularly at $\ell<100$, with a suppression of the excess around $\ell\sim70$, indicating that high-DM sources contribute appreciably to the measured signal. 
Taken together, these results suggest that the DM-dependent selection effects may affect the measured DM auto-power spectrum.
A more comprehensive assessment of selection effects, such as scattering, fluence-dependent selection, and other instrumental detection biases, will require further investigation in the future.

\section{Cosmological and Baryon Constraints}\label{parameters}
The measured DM angular power spectrum encodes the projected clustering of free electrons and thus serves as a powerful probe of cosmological and baryonic parameters. 
For a given set of parameters $\boldsymbol{\theta}$, we compute the theoretical DM angular power spectrum and construct the $\chi^2$ likelihood as
\begin{equation}
	\chi^2 (\boldsymbol{\theta}) = \left[
	C^{\rm obs}_{\ell} -
	\hat{C}^{\rm DM}_{\ell}(\boldsymbol{\theta})
	\right]^{\rm T} 	{\bf Cov}^{-1}
	\left[ C^{\rm obs}_{\ell}
	- \hat{C}^{\rm DM}_{\ell}(\boldsymbol{\theta}) 	\right],
\end{equation}
where $C^{\rm obs}_{\ell}$ is the vector of measured bandpowers, $\hat{C}^{\rm DM}_{\ell}(\boldsymbol{\theta})$ is the vector of theoretical bandpowers (Equations~(\ref{eq:cl_igm}), (\ref{eq:cl_DM}) and (\ref{eq:cl_noise})), 
and ${\bf Cov}$ is the covariance matrix.
We adopt the analytical Gaussian covariance for catalog-based pseudo-$C_\ell$ measurements developed by \cite{Wolz2026arXiv}. 
This method has been validated on simulations of sparse and noise-dominated FRB samples. 
Jackknife covariance estimation has also been employed in FRB-related two-point analyses \citep{Sharma2026ApJc}. 
A comparison between the Gaussian and jackknife covariance estimates for our data is presented in Appendix~\ref{app:jackknife}. 
We employ the Markov Chain Monte Carlo (MCMC) sampler \textsc{emcee} \citep{Foreman2013PASP} to sample the posterior distribution.

In this work, we consider two parameter combinations: the $\Omega_bh^2$--$H_0$ case and the $\Omega_bh^2$--$f_{\rm d}$ case. 
In both cases, we include two common free parameters, $\sigma_{\rm DM,\, res}$ and $\ell_{\rm loc}$. 
Here $\sigma_{\rm DM,\, res}$ represents a possible residual scale-independent DM contribution rather than the intrinsic scatter, since the dominant shot noise is already accounted for by the catalog-based estimator \citep{Wolz2025JCAP}. 
In the first case, the free cosmological parameters are the baryon density parameter $\Omega_bh^2$ and the Hubble constant $H_0$. 
We fix the matter density parameter $\Omega_m$ and adopt $f_{\rm d}=0.92$ for the cosmic LSS baryon fraction \citep{Wang2023ApJ, Connor2025NatAs, Liu2026ApJb}. 
This treats the DM angular power spectrum as a cosmological probe, testing its sensitivity to the cosmic baryon density and the expansion history. From this analysis,
we obtain $\Omega_bh^2=0.031^{+0.011}_{-0.019}$ and $H_0=74^{+20}_{-30}~{\rm km~s^{-1}~Mpc^{-1}}$, with $\sigma_{\rm DM,\, res}=19.3^{+8.7}_{-4.5}~{\rm pc~cm^{-3}}$ at the 68\% confidence level (CL). 
In the second case, the free cosmological parameters are $\Omega_bh^2$ and $f_{\rm d}$, while $\Omega_m$ and $H_0$ are fixed. 
This case focuses on the role of the DM angular power spectrum as a baryon tracer, since it is directly sensitive to the fraction of baryons in the LSS. 
We find $\Omega_bh^2=0.048^{+0.019}_{-0.035}$ and $f_{\rm d}=0.52^{+0.36}_{-0.24}$, with $\sigma_{\rm DM,\, res}=22^{+7}_{-4}~{\rm pc~cm^{-3}}$ at the $68\%$ CL. 
The angular scale of the localization uncertainty $\ell_{\rm loc}$ is unconstrained in both cases (see Figure~\ref{fig:posterior}). 

The one-dimensional marginalized posterior distributions and two-dimensional regions 
with $1-2\sigma$ contours corresponding to both parameter combinations are presented in  Figure~\ref{Fig3}. The corresponding parameter priors and constraints are summarized in Table~\ref{table1}.
The best-fit model for the $\Omega_{\mathrm{b}}h^2$--$H_0$ case is shown in Figure~\ref{Fig2}, overlaid on the measured DM angular power spectrum along with the residuals. 
Within the current uncertainties, the model provides a consistent description of the measured power spectrum. 
These results demonstrate that the angular power spectrum of FRB DMs can provide an effective probe of large-scale fluctuations in the ionized baryon distribution.

To further validate our methodology and assess the constraining power, Figure~\ref{Fig3} also shows the constraints obtained from $10^5$ mock FRB samples generated from the fiducial model (the construction of these mocks is described in Section~\ref{test}).
The mock posteriors successfully recover the fiducial input values and exhibit tighter contours, thereby clarifying the underlying parameter degeneracies.
In the $\Omega_{\mathrm{b}}h^2$--$H_0$ case, the posterior contour indicates that the DM power spectrum is sensitive to $\Omega_{\mathrm{b}}h^2$, while its dependence on $H_0$ becomes weak at high $H_0$. 
In the $\Omega_{\mathrm{b}}h^2$--$f_{\rm d}$ case, the degeneracy approximately follows $\Omega_{\mathrm{b}}h^2f_{\rm d}\simeq{\rm const}$, since both parameters control the overall amplitude of the projected DM fluctuations.
A comparison with the mock constraints suggests that the current measurement is predominantly limited by the FRB sample size. Nevertheless, 
these results show that the power-spectrum method can partially break these cosmological parameter degeneracies---unlike the mean ${\rm DM}_{\rm LSS}$--$z$ relation, where the parameters are completely degenerate, as indicated by the gray dot-dashed lines in Figure~\ref{Fig3} ($\Omega_{\mathrm{b}}h^2/H_0={\rm const}$ and $\Omega_{\mathrm{b}}h^2f_{\rm d}={\rm const}$).

\begin{deluxetable}{lccc}
	\centering
	\tablecaption{Parameter Constraints Derived from the Measured DM Power Spectrum for the Two Parameter Combinations Investigated. \label{table1}}
	\tablewidth{0pt}
	\tablehead{
		& Parameter & Prior & Estimation (68\% CL)
	}
	\startdata
	\hline
	\multirow{4}{*}{$\Omega_{\mathrm{b}}h^2$--$H_0$} & $\Omega_{b}h^2$ & (0, 0.1) &  $0.031^{+0.011}_{-0.019}$ \\
	& $ H_{0}/[\mathrm{km\; s^{-1}\; Mpc^{-1}}]$ & (10, 150) &  $74^{+20}_{-30}$  \\
	& $\sigma_{\rm DM,\, res}/[\mathrm{pc\; cm^{-3}}]$  & (0, 100) & $19.3^{+8.7}_{-4.5}$ \\
    & $\ell_{\rm loc}$ & 
    (2, 2000) & Unconstrained \\
	\hline
	\multirow{4}{*}{$\Omega_{\mathrm{b}}h^2$--$f_{\mathrm{d}}$} & $\Omega_{b}h^2$  &  (0, 0.1) & $0.048^{+0.019}_{-0.035}$ \\
	& $ f_{\mathrm{d}}$  & (0, 1) &  $0.52^{+0.36}_{-0.24}$\\
	& $\sigma_{\rm DM,\, res}/[\mathrm{pc\; cm^{-3}}]$ & (0, 100) &  $22^{+7}_{-4}$ \\
    & $\ell_{\rm loc}$ & 
    (2, 2000) & Unconstrained \\
	\enddata
    \tablecomments{The physical prior $\Omega_{b}h^2 \leq \Omega_{m}h^2$ is also adopted.}
\end{deluxetable}

\section{Sensitivity Tests and Systematics}\label{test}

Systematic uncertainties remain a major challenge for using FRB DMs as reliable cosmological probes \citep{Wang2025ApJ, Xu2025ApJ, Wang2026SSPMA, Sharma2026ApJb, Lemos2026arXiv}. 
Although the angular power spectrum is less susceptible to uncorrelated line-of-sight DM contributions than the conventional ${\rm DM}$--$z$ relation, the measured signal can still be influenced by assumptions regarding the FRB redshift distribution, the host-galaxy DM contribution, and the Milky Way foreground DM subtraction. 
It is therefore essential to test whether the parameter constraints inferred from the measured DM power spectrum are less sensitive to these systematic effects,
particularly as statistical precision improves in future FRB surveys.
In this section, we examine the impact of several representative sources of systematic uncertainty, including the assumed FRB redshift distribution, the host galaxy DM, the Milky Way halo DM, and the Galactic ISM electron-density model.
To this end, we first construct mock FRB samples from a fiducial DM model and then repeat the full analysis by varying one assumption at a time.
The resulting shifts in the inferred parameter constraints are used to quantify the systematic errors.

To generate the fiducial DM mock samples, we adopt the following procedure.
For each mock realization, FRB sky positions are randomly drawn over the full sky,
and the LSS DM fluctuation field is generated as a correlated full-sky random field using the \texttt{synfast} routine \citep{Gorski2005ApJ, Zonca2019JOSS}, with the input theoretical angular power spectrum $C^{\rm LSS-LSS}_\ell$ given by Eq.~(\ref{eq:cl_igm}).
The Milky Way ISM contribution, ${\rm DM}^{\rm MW}_{\rm ISM}$, is estimated from the NE2025 model \citep{Ocker2026ApJ},  
which is an updated Galactic electron-density model constrained by precise pulsar distances and scattering measurements of radio sources.
For the Milky Way halo contribution, we assume
that the probability distribution of $\mathrm{DM_{halo}^{MW}}$ follows a Gaussian distribution with mean $\mu_{\mathrm {halo}}=65$ $\mathrm{pc\;cm^{-3}}$ and standard deviation
$\sigma_{\mathrm {halo}}=15$ $\mathrm{pc\;cm^{-3}}$ \citep{Prochaska2019MNRAS, Wu2022MNRAS, Wang2023ApJ}:
\begin{equation}\label{eq:Phalo}
	P_{\rm halo}(\mathrm{DM_{halo}^{MW}})=\frac{1}{ \sqrt{2 \pi} \sigma_{\mathrm {halo}}} \exp \left[-\frac{\left(\mathrm{DM_{halo}^{MW}}-\mu_{\mathrm {halo}}\right)^2}{2 \sigma_{\mathrm {halo}}^{2}}\right]\;.
\end{equation}
For the host-galaxy DM, we adopt a log-normal distribution: 
\begin{equation}\label{eq:Phost}
	P_{\rm host}\left(\mathrm{DM}_{\mathrm{host}} \right)=\frac{1}{\sqrt{2 \pi}\mathrm{DM}_{\mathrm{host}} \sigma_{\mathrm {host}} } \exp \left[-\frac{\left(\ln \mathrm{DM}_{\mathrm{host}}-\mu_{\mathrm {host}}\right)^2}{2 \sigma_{\mathrm {host}}^{2}}\right]\;,
\end{equation}
where $\mu_{\mathrm{host}}$ and $\sigma_{\mathrm{host}}$ are the mean and standard deviation of $\ln \mathrm{DM}_{\mathrm {host}}$, respectively. 
Here we adopt $e^{\mu_{\mathrm{host}}}=90\, \mathrm{pc\;cm^{-3}}$ and $\sigma_{\mathrm{host}}=0.9$.
Both $\mathrm{DM_{halo}^{MW}}$ and $\mathrm{DM}_{\mathrm {host}}$ are independently drawn from their respective distributions.
For each mock catalog, the total mock DM is constructed by combining these DM components, i.e., $\mathrm{DM_{obs}^{mock}}=\mathrm{DM_{LSS}^{mock}}+\mathrm{DM_{halo}^{mock}}+{\rm DM}^{\rm mock}_{\rm ISM}+\mathrm{DM}_{\mathrm{host}}^{\rm mock}$.
We then measure the DM angular power spectrum from each mock sample following the same procedure as described in Section~\ref{Power measured}.
In the following subsections~\ref{redshift}--\ref{ISM}, we detail the alternative assumptions tested for the FRB redshift distributions, $\mathrm{DM}_{\mathrm{host}}$,  $\mathrm{DM_{halo}^{MW}}$ and ${\rm DM}^{\rm MW}_{\rm ISM}$. In all sensitivity tests except for the redshift-distribution test, we vary the alternative DM assumptions only in the generation of the mock data, while keeping the theoretical DM power spectrum fixed to the fiducial model. 
For each test, we generate mock samples containing $10^5$ bursts, a sample size representative of the statistical power expected from next-generation FRB surveys such as the Square Kilometre Array (SKA; 
\citealt{Zhang2023SCPMA}).
This larger mock sample is adopted to reduce the impact of statistical uncertainties and thereby to better reveal the systematic shifts that may become important for future high-precision measurements.

\subsection{Redshift Distribution}\label{redshift}

Although the DM power spectrum does not require precise redshift measurements for individual FRBs, 
it depends on the redshift distribution of the FRB sample through the projection weight function in Equation~(\ref{eq:dm_weight}).
The intrinsic FRB redshift distribution remains an open question and has been modeled using various approaches in the literature \citep{Qiang2021PhRvD, Zhang2021MNRAS, Zhang2022ApJ, Tang2023ChPhC, Gupta2025ApJ, Jia2026ApJ, Du2026arXiv}.
An inaccurate assumption regarding the redshift distribution can therefore introduce systematic biases in the inferred cosmological and baryonic parameters. It is thus essential to test the sensitivity of our results to the adopted form of $p_s(z)$.

In our fiducial model, we adopt the cutoff power-law (CPL) redshift distribution $p_s(z)\propto z^2\exp(-\alpha z)$ with $\alpha=3.5$. 
To assess the sensitivity of the DM power spectrum to the assumed FRB redshift distribution, we consider two alternative prescriptions. 
For the first alternative, we assume that the comoving number density of FRBs with intrinsic luminosity $L$ follows the Schechter function \citep{Schechter1976ApJ}:
\begin{equation}
	{\rm d} n(L) \propto \left(\frac{L}{L_{*}}\right)^{\alpha} \exp \left(-\frac{L}{L_{*}} \right) {\rm d} \left(\frac{L}{L_{*}}\right),
\end{equation}
where $\alpha$ is the power-law index and $L_*$ is the pivotal luminosity. 
The latter is calculated as $L_*=4\pi F_{\rm th}d_L^2(z_*)$, where $z_*$ denotes the corresponding pivotal redshift and $F_{\rm th}$ is the instrument flux threshold. 
For a flux-limited survey, the minimum detectable luminosity at redshift $z$ is $L_{\min}(z)=4\pi F_{\rm th} d_L^2 (z)$.
The corresponding FRB redshift distribution is then obtained from
\begin{equation}
	p_s(z) \propto \frac{{\rm d}V}{{\rm d} z} \int_{L_{\min}}^{\infty} {\rm d} n(L),
\end{equation}
where ${\rm d} V/ {\rm d}z$ is the differential comoving volume element.
We adopt $z_*=1.2$ and $\alpha=0.1$, following the values inferred from the FRB population model in the dispersion-galaxy cross-power spectrum measurement \citep{Wang2026PRL}.

\begin{figure}
    \centering
    \includegraphics[width=0.46\textwidth]{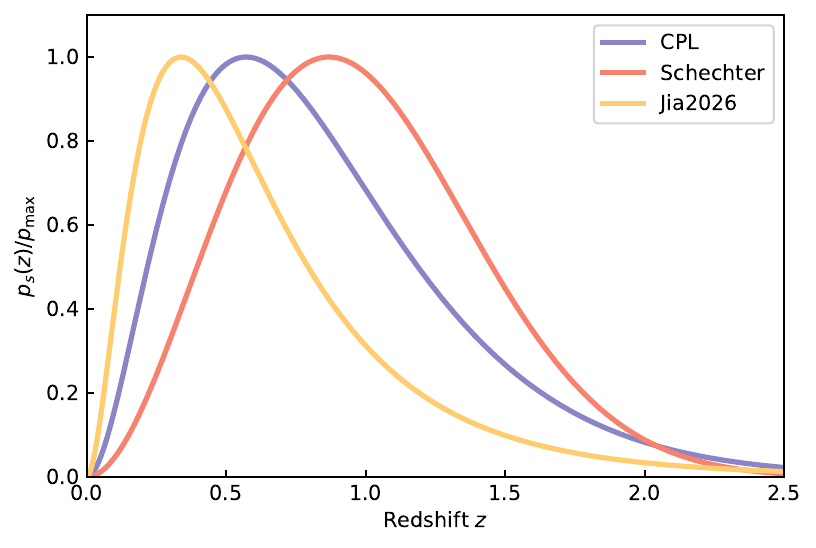}
    \caption{
    Comparison of the FRB redshift distributions adopted in this work.
    The CPL model is our fiducial prescription (purple line), while the Schechter model (\citet{Schechter1976ApJ}; orange line) and the distribution inferred by \citet{Jia2026ApJ} (yellow line) are used as
    alternatives in the sensitivity tests.
    Each distribution is normalized to its maximum.
    }
    \label{fig:redshift_distribution}
\end{figure}

The second alternative is the redshift distribution recently inferred from the second CHIME/FRB Catalog, which suggests a rapid decline of the FRB formation rate relative to the star-formation rate, described by a simple power-law form \citep{Jia2026ApJ}, hereafter referred to as the Jia2026 model. 
The redshift distribution is modeled as
\begin{equation}
    p_s(z) \propto \frac{\rho_{\rm FRB}(z)}{1+z} \frac{{\rm d}V}{{\rm d} z},
\end{equation}
where $\rho_{\rm FRB}(z) \propto (1+z)^{-5.38} $ denotes the comoving FRB formation rate.
In all cases, the redshift distributions are normalized over the redshift range considered in our analysis using the same normalization procedure. 
Figure~\ref{fig:redshift_distribution} compares the three FRB redshift distributions considered in our analysis. 
The three models present different redshift-dependent profiles:
the Jia2026 model peaks at the lowest redshift, the CPL distribution at an intermediate redshift, and the Schechter distribution at the highest redshift. 
These distinct redshift distributions therefore allow us to assess the sensitivity of the DM angular power spectrum to uncertainties in the assumed FRB redshift distribution.

\subsection{Host Galaxy}\label{host}

The host-galaxy DM contribution, $\mathrm{DM}_{\mathrm{host}}$, constitutes a major source of systematic uncertainty in cosmological applications of FRBs \citep{Zhang2020ApJ, Mo2023MNRAS, Reischke2025OJAp, Li2025ApJ, Xu2025ApJ, Sang2025MNRAS, Jia2026PhLB, MasRibas2026}.
This uncertainty stems from the intrinsic diversity of host galaxy properties, the wide range of local environments and the varying physical locations of FRBs within their hosts, all of which make ${\rm DM}_{\rm host}$ difficult to estimate accurately. 
A log-normal distribution, as given in Equation~(\ref{eq:Phost}), is commonly adopted to model the statistical behavior of ${\rm DM}_{\rm host}$ \citep{Macquart2020Natur, Zhang2020ApJ}. 
However, the mean and scatter of this distribution remain poorly constrained and differ substantially across independent studies.
To evaluate the sensitivity of our results to this uncertainty, we consider two alternative cases: a high-$\sigma_{\rm host}$ case, with $e^{\mu_{\mathrm{host}}}=90\; \mathrm{pc\;cm^{-3}}$ and $\sigma_{\mathrm{host}}=1.2$, representing a broader scatter; and a high-$\mu_{\rm host}$ case, with $e^{\mu_{\mathrm{host}}}=150\; \mathrm{pc\;cm^{-3}}$ and $\sigma_{\mathrm{host}}=0.9$, representing a larger mean host DM.

\subsection{Milky Way Halo}\label{halo}

The Milky Way halo contribution, ${\rm DM}^{\rm MW}_{\rm halo}$, represents another source of foreground uncertainty in DM measurements. 
Various models and empirical estimates have been proposed for this component \citep{Prochaska2019MNRAS, Keating2020MNRAS, Das2021MNRAS, Ravi2025AJ, Wang2025Univ, Liu2026ApJ, Zhang2026arXiv, Wang2026RAA}. 
In our fiducial analysis, we model ${\rm DM}^{\rm MW}_{\rm halo}$ as an isotropic Gaussian distribution, as given in Eq.~(\ref{eq:Phalo}). While this prescription captures the expected mean level of the Galactic halo contribution, it neglects potential direction-dependent variations, which may introduce systematic biases in the measured DM power spectrum.

To assess the impact of such anisotropies, we consider two alternative anisotropic ${\rm DM}^{\rm MW}_{\rm halo}$ models. 
First, we adopt the analytic Galactic halo model of \citet{Yamasaki2020ApJ} (YT2020 model), which uses diffuse X-ray observations to construct a Milky Way halo DM model including both a spherical component and a disk-like nonspherical component. 
This model is parameterized as a seventh-order polynomial, ${\rm DM}^{\rm MW}_{\rm halo}=\sum_{i,j} c_{ij}|l|^i|b|^j$, where $l$ and $b$ are the Galactic longitude and latitude, respectively, and $c_{ij}$ are the fitted coefficients.
Second, we consider the angular power spectrum of ${\rm DM}^{\rm MW}_{\rm halo}$ presented in Figure~2 of \citet{Huang2025MNRAS} (hereafter Huang25 model).
That spectrum was derived from their local-universe DM model, in which the Milky Way halo DM contribution is obtained from the \textsc{Hestia} hydrodynamic simulations \citep{Libeskind2020MNRAS}. 
For implementation, we parameterize the halo power spectrum as
\begin{equation}
	\ell(\ell+1) C_\ell^{\rm halo} =
	\begin{cases}
		C_0, & \ell < \ell_0, \\
		C_0(\ell/\ell_0)^{-\beta}, & \ell \geq \ell_0,
	\end{cases}
\end{equation}
with $C_0=500$ ${\rm (pc\, cm^{-3})^2}$, $\ell_0=15$, and $\beta=1$.
This parameterized spectrum is then used to generate anisotropic ${\rm DM}^{\rm MW}_{\rm halo}$ in the mock catalogs.

\subsection{Milky Way ISM}\label{ISM}

Several Galactic electron density models have been developed to estimate ${\rm DM}^{\rm MW}_{\rm ISM}$, including the NE2001 model \citep{Cordes2002}, the YMW16 model \citep{Yao2017ApJ}, and the recently updated NE2025 model \citep{Ocker2026ApJ}.
In our fiducial analysis, we adopt NE2025 as the reference model and treat it as an unbiased estimator of the Galactic ISM contribution.
To test the sensitivity of the power-spectrum method to the choice of the Galactic electron-density model, we repeat the analysis using NE2001 and YMW16 in place of NE2025.
Specifically, for each FRB, we compute the difference between the NE2025 model and another ISM model, 
\begin{equation}
	\Delta {\rm DM}^{\rm MW}_{\rm ISM}
	=	{\rm DM}_{\rm NE2025}	-
	{\rm DM}_{\rm NE2001(YMW16)}. 
	\label{eq:delta_dm_ism}
\end{equation}
We then update the foreground-corrected DM for each burst by applying this difference and repeat the power-spectrum measurement and subsequent parameter inference.

\subsection{Test Results}

\begin{figure*}
	\centering
	\includegraphics[width=0.49\textwidth]{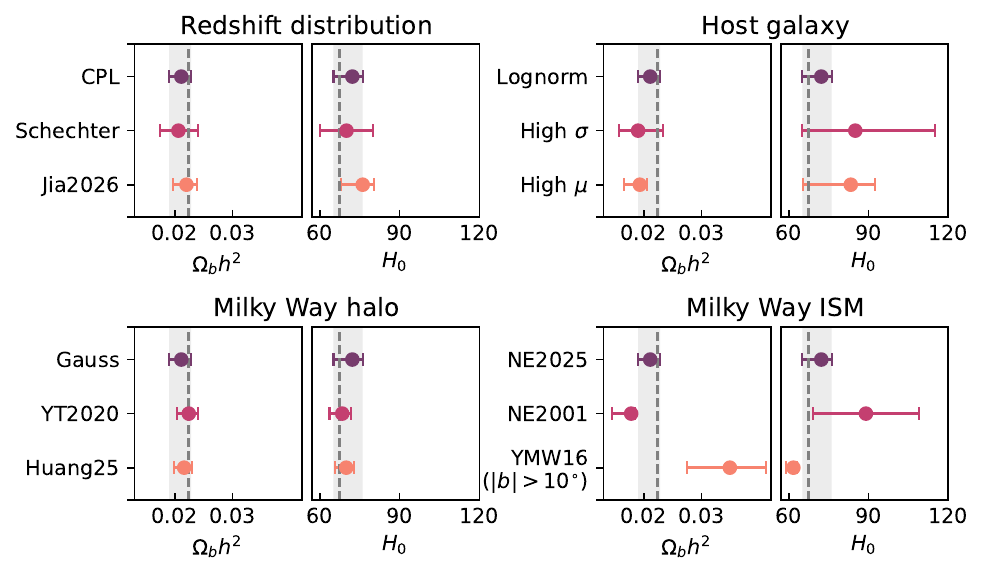}
	\includegraphics[width=0.49\textwidth]{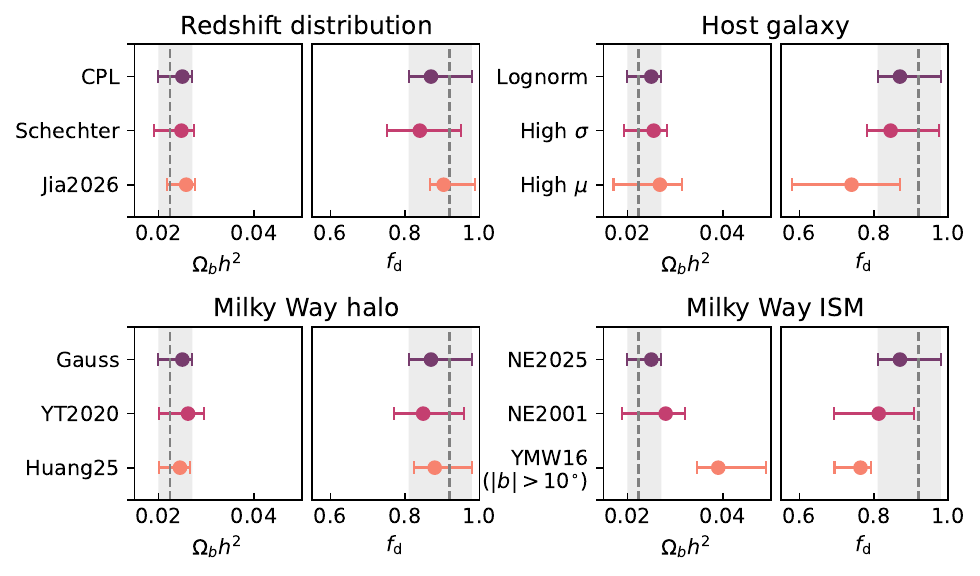}
	\vskip-0.1in
	\caption{Sensitivity tests for parameter constraints under different modeling assumptions.
		The left and right panels show the results for the $\Omega_{\rm b}h^2$--$H_0$ and $\Omega_{\rm b}h^2$--$f_{\rm d}$ cases, respectively. 
		In each panel, the points with error bars represent the one-dimensional marginalized constraints obtained by varying one assumption at a time, including the FRB redshift distribution, the host-galaxy DM model, the Milky Way halo model, and the Milky Way ISM model.
        The gray dashed lines indicate the fiducial input values, and the gray shaded regions denote the 1$\sigma$ credible intervals from the fiducial analysis.}
	\label{Fig5}
\end{figure*}

To quantify the deviation of the constraints obtained under each alternative assumption from those of the fiducial model, we define the bias metric
\begin{equation}\label{Delta}
	\Delta = \frac{|x_{\rm test}-x_{\rm fid}|}
	{\sqrt{\sigma_{\rm test}^2+\sigma_{\rm fid}^2}},
\end{equation}
where $x_{\rm test}$ and $x_{\rm fid}$ denote the best-fit values from a given sensitivity test and the fiducial analysis, respectively, and $\sigma_{\rm test}$ and $\sigma_{\rm fid}$ are 
their corresponding $1\sigma$ uncertainties.
The results of all sensitivity tests are shown in Figure~\ref{Fig5} and summarized in Table~\ref{tab:test}. 

In the $\Omega_{\rm b}h^2$--$H_0$ case, the inferred values of $\Omega_{\rm b}h^2$ are generally lower than the fiducial values across most tests.
This behavior arises from the weak sensitivity of the DM angular power spectrum to large values of $H_0$, which broadens the posterior distribution in the high-$H_0$ region and shifts the one-dimensional marginalized distribution of $\Omega_{\rm b}h^2$ toward lower values, as also seen in Figure~\ref{Fig3}.
The host-galaxy DM contribution has only a weak impact on the constraints. 
Because ${\rm DM}_{\rm host}$ is largely uncorrelated across different FRB sightlines, it contributes primarily to the stochastic DM noise rather than to the correlated power spectrum signal. 
By contrast, the Galactic foreground components, particularly the anisotropic Milky Way ISM, have a more noticeable effect.
This impact is especially evident in the choice of the Galactic ISM model:
the YMW16 model differs significantly from both NE2025 and NE2001, and the resulting parameter shifts remain significant after removing FRBs located at low Galactic latitudes ($|b|<10^\circ$) and increasing the minimum multipole to $\ell_{\min}=40$.

Overall, these tests demonstrate that the DM angular power spectrum method is less sensitive to scale-independent DM components, such as the stochastic host‑galaxy contribution, but remains sensitive to foreground terms that possess anisotropic spatial structures.
Galactic foreground modeling, especially of the anisotropic Milky Way ISM, therefore constitutes an important source of systematic uncertainty for current and future measurements.
Compared with the mean ${\rm DM}_{\rm LSS}$--$z$ relation, the power-spectrum approach mitigates a larger fraction of the systematic uncertainties associated with DM component modeling, making it a more reliable cosmological probe and baryon tracer. Nevertheless, accurate and physically motivated modeling of anisotropic Galactic foregrounds will be essential for fully exploiting the statistical power of larger FRB samples expected from upcoming surveys.

\definecolor{Purple}{RGB}{175, 169, 249}
\providecommand{\deltacell}[2]{%
	\cellcolor{Purple!#1!white}%
	\makebox[0.8cm][c]{#2}%
}
\begin{table*}
	\centering
	\caption{Summary of Sensitivity Tests for Cosmological and Baryonic Constraints.}
	\label{tab:test}
	\begin{tabular}{c c c c c c c c c c c}
		\hline
		\hline
		\multirow{2}{*}{ } & \multirow{2}{*}{ } &
		\multicolumn{4}{c}{$\Omega_{\mathrm{b}}h^2$--$H_0$} & &
		\multicolumn{4}{c}{$\Omega_{\mathrm{b}}h^2$--$f_{\mathrm{d}}$} \\
		\cline{3-6} \cline{8-11}
		& & $\Omega_{\mathrm{b}}h^2$ & $\Delta$ & $H_0/[\mathrm{km\; s^{-1}\; Mpc^{-1}}]$ & $\Delta$ & & $\Omega_{\mathrm{b}}h^2$ & $\Delta$ & $f_{\mathrm{d}}$ & $\Delta$ \\
		\hline
		& Fiducial  & $0.0211^{+0.0017}_{-0.0021}$ & $\cdots$ & $72.2^{+3.9}_{-7.1}$ & $\cdots$ & & $0.0250^{+0.0021}_{-0.0050}$ & $\cdots$ & $0.870^{+0.060}_{-0.110}$ & $\cdots$ \\
		\hline
		\multirow{2}{*}{Redshift distribution} & Schechter & $0.0206^{+0.0034}_{-0.0032}$  & \deltacell{4}{0.13} & $70.0^{+10}_{-10}$       & \deltacell{6}{0.18} & & $0.0248^{+0.0026}_{-0.0057}$  & \deltacell{1}{0.04} & $0.840^{+0.110}_{-0.089}$    & \deltacell{6}{0.19} \\
		& Jia2026 & $0.0220^{+0.0018}_{-0.0023}$  & \deltacell{10}{0.31}  & $76.1^{+4.4}_{-8.1}$       & \deltacell{14}{0.43} & & $0.0258^{+0.0018}_{-0.0041}$  & \deltacell{6}{0.17} & $0.904^{+0.085}_{-0.037}$  & \deltacell{16}{0.48} \\
		\hline
		\multirow{2}{*}{Host galaxy} & High $\sigma$ & $0.0190^{+0.0043}_{-0.0033}$  & \deltacell{14}{0.44} & $85.0^{+20}_{-20}$           & \deltacell{21}{0.63} & & $0.0255^{+0.0027}_{-0.0062}$  &  \deltacell{3}{0.08} & $0.845^{+0.130}_{-0.064}$    & \deltacell{5}{0.15} \\
		& High $\mu$ & $0.0193^{+0.0012}_{-0.0027}$  & \deltacell{24}{0.74} & $83.3^{+9.2}_{-18}$           & \deltacell{20}{0.60} & & $0.0268^{+0.0046}_{-0.0097}$  & \deltacell{6}{0.18} & $0.740^{+0.130}_{-0.160}$    & \deltacell{25}{0.76} \\
		\hline
		\multirow{2}{*}{Milky Way halo} & YT2020 & $0.0224^{+0.0016}_{-0.0020}$  & \deltacell{17}{0.50} & $68.4^{+3.2}_{-4.8}$          & \deltacell{16}{0.49} & & $0.0262^{+0.0034}_{-0.0060}$  & \deltacell{6}{0.19} & $0.849^{+0.110}_{-0.079}$    & \deltacell{4}{0.13} \\
		& Huang25 & $0.0216^{+0.0014}_{-0.0017}$  & \deltacell{7}{0.21} & $69.9^{+2.9}_{-4.3}$         & \deltacell{10}{0.30} & & $0.0245^{+0.0021}_{-0.0044}$  & \deltacell{3}{0.09}  & $0.880^{+0.100}_{-0.056}$  & \deltacell{4}{0.12} \\
		\hline
		\multirow{2}{*}{Milky Way ISM} & NE2001 & $0.0178^{+0.0006}_{-0.0034}$  & \deltacell{50}{1.51} & $89.0^{+20}_{-20}$        & \deltacell{28}{0.82} & & $0.0280^{+0.0041}_{-0.0091}$  & \deltacell{11}{0.32} & $0.813^{+0.095}_{-0.120}$  & \deltacell{13}{0.39} \\
		& YMW16 & $0.0349^{+0.0063}_{-0.0074}$  & \deltacell{60}{1.82} & $61.7^{+1.3}_{-2.8}$       & \deltacell{48}{1.45} & & $0.0390^{+0.0100}_{-0.0044}$ & \deltacell{97}{2.9} & $0.764^{+0.029}_{-0.070}$  & \deltacell{31}{0.93} \\
		\hline
	\end{tabular}
	\tablecomments{The table summarises the parameter constraints obtained under variations in the  assumed FRB redshift distribution, host-galaxy DM, Milky Way halo DM, and Milky Way ISM model.
		The columns are divided into two blocks, corresponding to the $\Omega_{\rm b}h^2$--$H_0$ and $\Omega_{\rm b}h^2$--$f_{\rm d}$ fitting cases, respectively. For each tested scenario, 
		$\Delta$ denotes the normalized deviation from the fiducial result, as defined in Equation~(\ref{Delta}).
		The purple shades indicate the magnitude of the deviation $\Delta$, with darker shades corresponding to larger values.
	}
\end{table*}

\section{Conclusion and Discussion}
\label{Sec:Conclusion}

In this work, we have presented the first measurement of the angular auto-power spectrum of FRB DMs, using 3455 apparently non-repeating bursts from the CHIME/FRB Catalog 2.
After subtracting the Milky Way ISM foreground contribution, ${\rm DM}^{\rm MW}_{\rm ISM}$, we construct a residual-DM field and detect its auto-correlation signal at $\sim 2.8 \sigma$ significance.
We fit the measured bandpowers with the theoretical power spectrum to constrain cosmological and baryonic parameters, and we further test the sensitivity of the results against several systematic uncertainties related to the assumed FRB redshift distribution and the modeling of the various DM components.
Our main conclusions are summarized as follows:
\begin{enumerate}
	\item  
    The detected angular auto-correlation signal provides evidence that the FRB DM field contains statistically significant two-point correlation information.
	The measured DM angular power spectrum thus constitutes a novel and powerful observable tool for probing cosmic electron-density fluctuations.
	
	\item By fitting the measured bandpowers to the theoretical spectrum, we obtain constraints on two parameter combinations. 
	The $\Omega_{\rm b}h^2$--$H_0$ case probes the cosmic baryon density and the expansion rate of the Universe, while the $\Omega_{\rm b}h^2$--$f_{\rm d}$ case traces the cosmic baryon density and the baryon fraction in the cosmic LSS. 
	Although the current constraints are still limited by the available FRB sample size, they clearly demonstrate the potential of the DM power spectrum as both a competitive cosmological probe and an effective baryon tracer. 
	
	\item Our sensitivity tests indicate that the power-spectrum method is largely insensitive to uncorrelated DM components, such as the stochastic host-galaxy contribution ${\rm DM}_{\rm host}$. 
	Therefore, this method can reduce the impact of systematic uncertainties associated with the difficult-to-estimate host-galaxy DM. 
	In contrast, anisotropic Galactic foregrounds have a more noticeable effect on the inferred parameters. 
	Uncertainties in the modeling of ${\rm DM}^{\rm MW}_{\rm ISM}$ therefore remain important sources of systematic uncertainty. 
	Accurate and physically motivated modeling of Galactic foregrounds will be essential for fully exploiting the statistical power of future, larger FRB samples.
	
	\item Compared with the traditional approach based directly on the ${\rm DM}_{\rm LSS}$--$z$ relation, the DM angular power spectrum offers two advantages. 
	First, it does not require precise redshift measurements for individual bursts; instead, it depends only on the overall redshift distribution of the sample. 
	This feature allows the inclusion of large FRB samples lacking host-galaxy identifications.
	Second,  the power-spectrum method can partially break the parameter degeneracies inherent in the ${\rm DM}_{\rm LSS}$--$z$ relation, enabling the measured bandpowers to be jointly analyzed in two parameter planes: $\Omega_{b}h^2$--$H_0$ and $\Omega_{b}h^2$--$f_{\rm d}$.
	
\end{enumerate}

Although the current constraints are limited by the FRB sample size and incomplete sky coverage, our results nevertheless demonstrate that the FRB DM angular power spectrum encodes valuable information about cosmic baryon fluctuations. 
Future surveys, such as the Deep Synoptic Array (DSA-2000; \citealt{Hallinan2019BAAS}) and the SKA \citep{Dewdney2009IEEEP}, are expected to deliver substantially larger FRB samples with wider sky coverage \citep{Hashimoto2020MNRAS, Zhang2023SCPMA}. 
These advancements will significantly enhance the sensitivity of the power-spectrum method and will establish the FRB DM angular power spectrum as a powerful probe of cosmology and the large-scale distribution of ionized baryons.

\section*{Acknowledgments}
This work is supported by the National Natural Science Foundation of China (grant Nos. 12422307 and 12373053), 
the Strategic Priority Research Program of the Chinese Academy of Sciences (grant No. XDB0550400), the National Key R\&D 
Program of China (grant Nos. 2024YFA1611704 and 2024YFA1611700), the China Manned Space Program (grant No. CMS-CSST-2025-A01), 
and China Postdoctoral Science Foundation (grant No. 2025M783235).

\appendix

\section{Impact of Baryonic Feedback Parameter}
\label{app:baryon}

\begin{figure}
	\centering
	\includegraphics[width=0.46\textwidth]{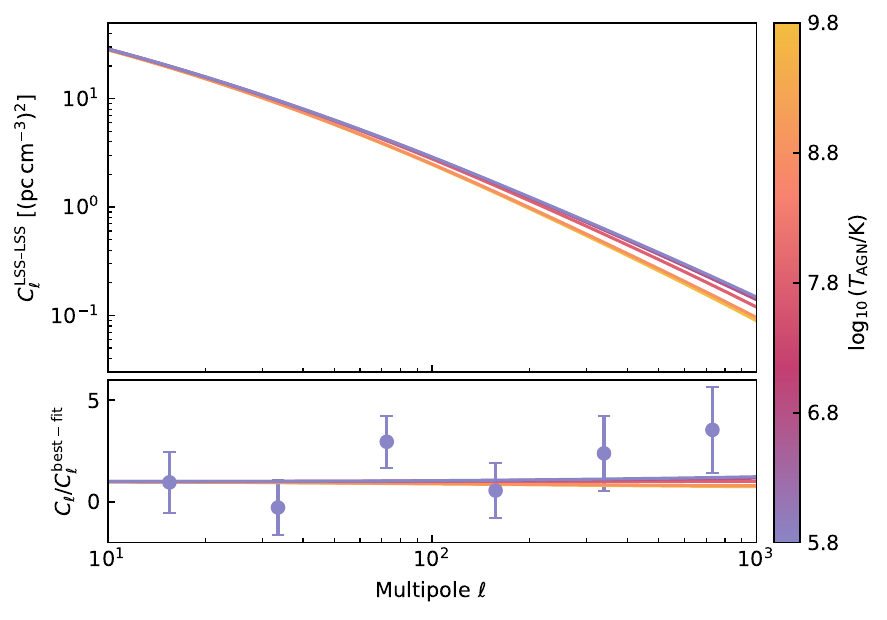}
	\caption{
        Impact of the AGN feedback parameter $T_{\rm AGN}$ on the ${\rm DM_{LSS}}$ angular power spectrum $C_\ell^{\rm LSS-LSS}$.
        The upper panel shows the theoretical curves of $C_\ell^{\rm LSS-LSS}$ for different values of $\log_{10}(T_{\rm AGN}/{\rm K})$, while the lower panel presents the ratio of the theoretical curves and the measured values relative to the best-fit model with $\log_{10}(T_{\rm AGN}/{\rm K})=7.8$. 
        The shot-noise contribution has been subtracted from the data.
        The baryonic feedback effect becomes increasingly significant towards higher multipoles but remains much smaller than the current data errors.
		}
	\label{fig:TAGN}
\end{figure}

Baryonic feedback can modify the matter distribution on nonlinear scales and therefore affect the DM angular power spectrum, particularly at high multipoles $\ell$. 
To quantify this effect, we vary the AGN feedback parameter $T_{\rm AGN}$ over the range $\log_{10}(T_{\rm AGN}/{\rm K})\in [5.8, 9.8]$ and calculate the corresponding ${\rm DM_{LSS}}$ power spectra.
Figure~\ref{fig:TAGN} shows the dependence of the ${\rm DM_{LSS}}$ power spectrum on $T_{\rm AGN}$. 
The feedback effect becomes increasingly significant towards higher multipoles, reaching a maximum deviation of $\sim20\%$ at $\ell\sim1000$ relative to our best-fit model with $\log_{10}(T_{\rm AGN}/{\rm K})=7.8$. 
However, the current measurement uncertainties are relatively large, so that the variations induced by $T_{\rm AGN}$ remain subdominant for the present data.
We therefore fix the value of $T_{\rm AGN}$ in our analysis.

\section{Covariance Estimation}
\label{app:jackknife}

\subsection{Jackknife covariance}

The jackknife method, originally developed in the 1950s as a general statistical tool \citep{Quenouille1956}, has since been widely adopted for estimating two-point clustering statistics in galaxy surveys \citep{Norberg2009MNRAS, Friedrich2016MNRAS}, and has also been applied to covariance estimation in DM-related power spectrum analyses \citep{Sharma2026ApJc}.
In this analysis, we partition the CHIME/FRB sky coverage into $N_{\rm patch}$ equal-area regions defined according to the angular positions of the FRBs. 
In the $k$-th jackknife realization, all bursts located in region $k$ are removed, and the DM angular power spectrum is recomputed from the remaining sample using the same estimator and multipole binning as in the fiducial analysis.
The jackknife covariance matrix is then estimated as
\begin{equation}\label{eq:jackknife_cov}
	{\rm Cov}_{ij} = \frac{N_{\rm patch}-1}{N_{\rm patch}}
	\sum_{k=1}^{N_{\rm patch}}
	\left(\hat C_i^{(k)}-\bar C_i\right)
	\left(\hat C_{j}^{(k)}-\bar C_{j}\right).
\end{equation}
where $\hat C_i^{(k)}$ is the measured bandpower in multipole bin $i$ for the $k$-th jackknife realization, and $\bar C_i$ is the mean bandpower averaged over all jackknife realizations. 
We adopt $N_{\rm patch}=118$ in this work.
The $1\sigma$ uncertainty for the $i$-th bandpower is given by the square root of the corresponding diagonal element of the covariance matrix,
\begin{equation}
	\sigma_i = \sqrt{{\rm Cov}_{ii}}
	=	\left[ 	\frac{N_{\rm patch}-1}{N_{\rm patch}}
	\sum_{k=1}^{N_{\rm patch}}
	\left(\hat C_i^{(k)}-\bar C_i\right)^2 	\right]^{1/2}.
\end{equation}

Since the covariance matrix ${\bf Cov}$ is estimated from a finite number of jackknife patches, its inverse is subject to a finite-sample bias. We therefore apply the Hartlap correction \citep{Hartlap2007A&A} when using the inverse covariance matrix in the parameter inference,
\begin{equation}
\mathbf{Cov}^{-1}_{H}
= \frac{N_{\rm patch}-N_{\rm data}-2}
     {N_{\rm patch}-1}
\mathbf{Cov}^{-1},
\label{eq:hartlap}
\end{equation}
where $N_{\rm data}$ is the length of the bandpower data vector. 
This correction reduces the finite-sample bias in the inverse covariance arising from the limited number of jackknife patches.

In addition, the uncertainty in the estimated covariance matrix also propagates into the inferred parameter uncertainties \citep{Dodelson2013PhRvD}. 
Following \citet{Percival2014MN}, the parameter covariance is corrected as
\begin{equation}
\sigma^2_{p,\star}=m\sigma^2_p,
\end{equation}
where
\begin{equation}
m=\frac{1+B(N_{\rm data}-N_{\rm par})}
{1+A+B(N_{\rm par}+1)},
\end{equation}
with
\begin{equation}
A=\frac{2}{(N_{\rm patch}-N_{\rm data}-1)(N_{\rm patch}-N_{\rm data}-4)},
\end{equation}
\begin{equation}
B=\frac{N_{\rm patch}-N_{\rm data}-2}
{(N_{\rm patch}-N_{\rm data}-1)(N_{\rm patch}-N_{\rm data}-4)}.
\end{equation}
Here, we use $N_{\rm patch}=118$, $N_{\rm data}=6$, and $N_{\rm par}=4$.

\subsection{Covariance comparison}

\begin{figure*}
	\centering
	\includegraphics[width=0.8\textwidth]{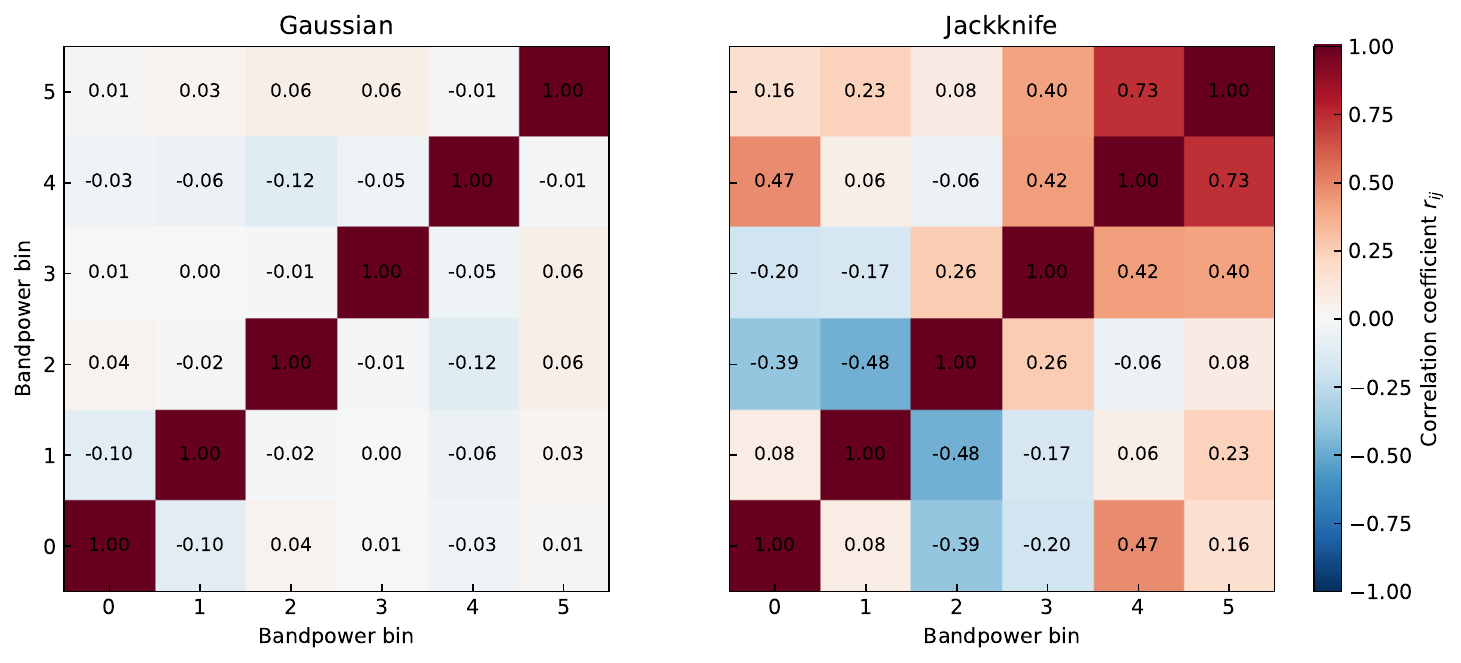}
	\caption{
        Correlation coefficient matrices of the measured DM angular power spectrum, derived from the Gaussian covariance estimate (left) and jackknife covariance estimates with $N_{\rm patch}=118$ (right). 
		Each matrix element shows the correlation coefficient $r_{ij}$ between the six logarithmically spaced multipole bins. 
        The presence of non-zero off-diagonal elements indicates that the bandpowers are not independent, motivating the use of the full covariance matrix in the likelihood analysis.
	}
	\label{Correlation}
\end{figure*}

To quantify the correlations between measurements at different multipoles, we compute the corresponding correlation coefficient matrix,
\begin{equation}
	r_{ij} = \frac{{\rm Cov}_{ij}}
	{\sqrt{{\rm Cov}_{ii}{\rm Cov}_{jj}}}.
\end{equation}
The correlation coefficient matrices derived from the Gaussian covariance estimate and the jackknife estimates with $N_{\rm patch}=118$ are shown in Figure~\ref{Correlation}. 
The Gaussian covariance exhibits weak off-diagonal correlations, whereas the jackknife covariance shows stronger correlations between different multipole bins. 
For sparse catalogs, the shot-noise component of the discrete angular mask can induce strong coupling between widely separated angular scales \citep{Wolz2025JCAP}. 
In addition, this difference may partly arise from the noise introduced by a finite number of jackknife resamplings \citep{Favole2021MNRAS}.

\begin{figure}
	\centering
	\includegraphics[width=0.46\textwidth]{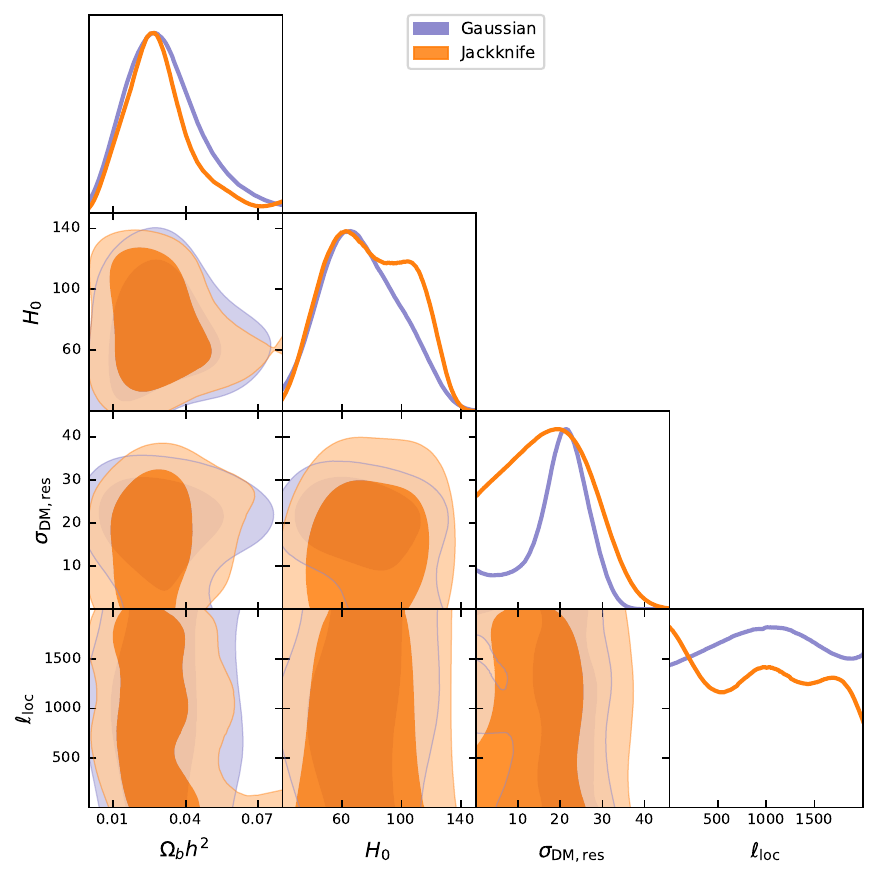}
	\caption{
    The posterior probability distributions obtained using the Gaussian (purple) and jackknife covariances (orange).
    The dark and light shaded regions indicate the 68\% and 95\% confidence intervals, respectively.
	}
	\label{fig:posterior}
\end{figure}

To assess the impact of these differences on the cosmological inference, we repeat the parameter analysis using the Gaussian and jackknife covariances. 
As shown in Figure~\ref{fig:posterior}, the posterior distributions obtained with the two covariances are broadly consistent. 
Despite the differences between the two covariance estimates, their impact on the inferred parameter constraints is negligible.

\section{Significance Test}
\label{app:significance}

To quantify the preference of the model over the null hypothesis, we adopt the $\Delta\chi^2$ test. 
The null hypothesis assumes that the measured power spectrum is a zero signal ($C_{\ell}=0$). We define
\begin{equation}
\Delta\chi^2
\equiv \chi^2_0-\chi^2_{\rm min},
\end{equation}
where $\chi^2_0$ is calculated between the null model and the measured bandpowers, 
and $\chi^2_{\rm min}$ is the minimum value obtained between the best-fit signal model and the data.
The effective number of degrees of freedom is estimated using the Bayesian model dimensionality \citep{Handley2019PhRvD},
\begin{equation}
d_M=2\left[\langle (\ln\mathcal{L})^2\rangle
-\langle\ln\mathcal{L}\rangle^2\right],
\end{equation}
where $\mathcal{L}$ is the likelihood function, with $\ln \mathcal{L}=-\chi^2(\boldsymbol{\theta})/2+{\rm const}$, and the expectation values are evaluated over the posterior distribution. 
This quantity accounts for parameter degeneracies and provides an effective dimensionality of the model.
For our analysis, we obtain $\Delta\chi^2 \simeq 8.6$ with an effective dimensionality of $d_M \simeq 1.2$, corresponding to a significance level of approximately $2.8 \sigma$.

\section{Multipole Binning Test}
\label{app:binning}

\begin{figure}
	\centering
	\includegraphics[width=0.46\textwidth]{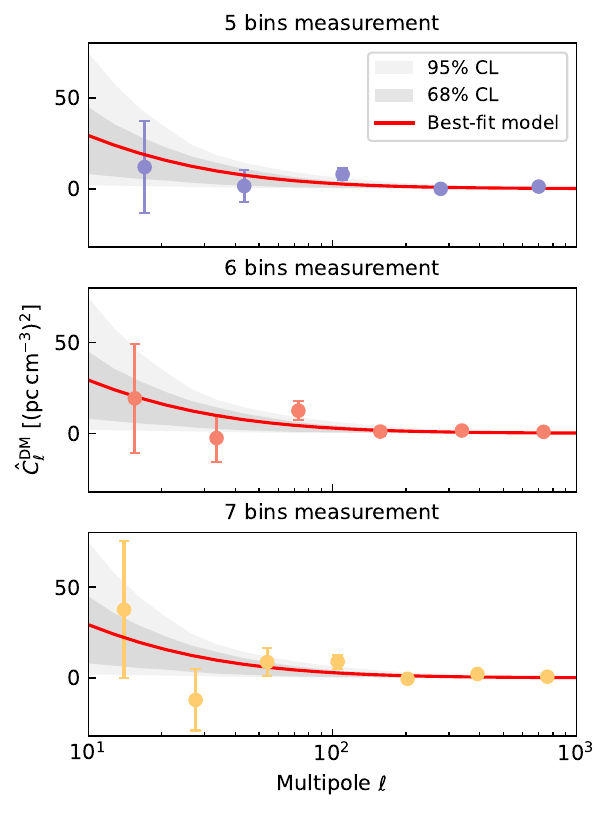}
	\caption{
        Test of the measured DM angular power spectrum with different choices of multipole binning.
        The results with 5 bins (purple points), 6 bins (orange points) and 7 bins (yellow points) are shown.
        The red curves denote the best-fit models, while the dark and light gray regions represent the $68\%$ and $95\%$ CLs, respectively.
        The consistency among different binning measurements indicates that the measured signal is not driven by a choice of multipole binning.
		}
	\label{fig:binning}
\end{figure}

To test the sensitivity of the measured DM angular power spectrum to the choice of multipole binning, we repeat the analysis using different numbers of multipole bins. 
In addition to our fiducial choice of 6 bins, we consider alternative binning schemes with 5 and 7 bins. 
The measurements are shown in Figure~\ref{fig:binning}. 
We quantify the impact on the signal significance by calculating the difference in $\chi^2$ between the null model and the best-fit model. 
For 5, 6, and 7 bins, we obtain $\Delta \chi^2 = 8.9$, 8.6, and 8.4 respectively.
The small variation in $\Delta\chi^2$ indicates that the signal detection is insensitive to the number of multipole bins.

\section{Null Test}
\label{app:null}

\begin{figure}
	\centering
	\includegraphics[width=0.46\textwidth]{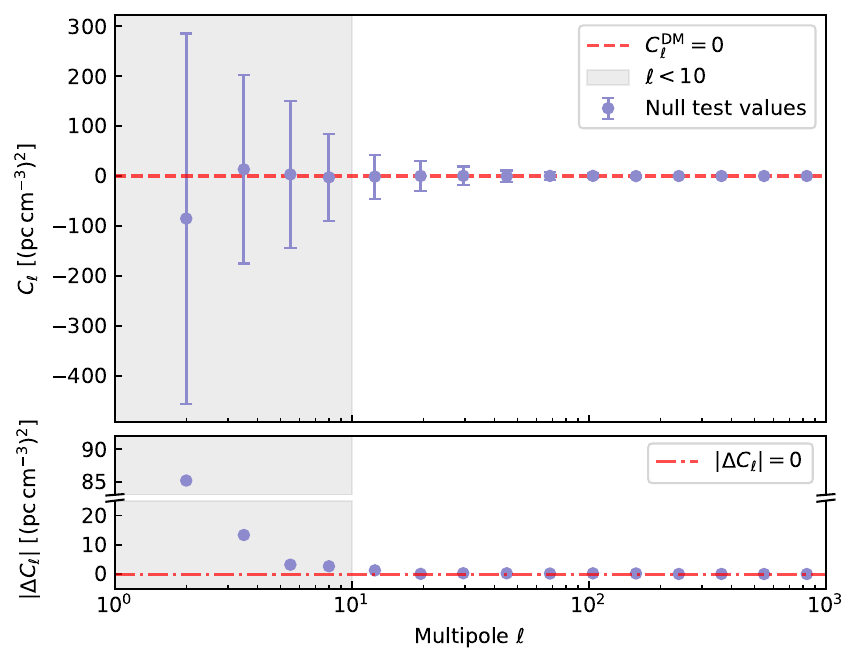}
	\caption{
        Null test of the DM angular power spectrum.
        Upper panel: the null power spectra measured from randomized FRB catalogs that preserve the survey selection function and sky coverage.
        The dashed red line indicates the null expectation $C_\ell^{\rm DM}=0$.
        The gray shaded region denotes the low-multipole range $\ell<10$.
        Lower panel: the deviation from the null hypothesis, $|\Delta C_{\ell}|$.
        The dot-dashed red line indicates $|\Delta C_{\ell}|=0$ ${\rm (pc\,cm^{-3})^2}$.
		}
	\label{fig:null}
\end{figure}

Large-scale angular modes can be affected by survey geometry and non-uniform sky coverage of FRBs \citep{Ribeiro2026arXivb}.
To assess whether these effects introduce interfering signals in the measured DM angular power spectrum, we perform a null test in which the FRB sky positions are kept fixed while the residual DM values are randomly reassigned across these positions.
For each mock catalog, the DM angular power spectrum is recomputed using the same estimator as in the fiducial analysis. 
From 1000 randomized mock catalogs, we construct the mean and uncertainty for each of the 15 multipole bins.
We then quantify the deviation from the null hypothesis as $|\Delta C_{\ell}|=|C_{\ell}-C_{\ell}^{\rm null}|$.
The resulting null power spectra are shown in Figure~\ref{fig:null}. 
The multipoles $\ell<10$ show large scatter and deviations from zero, as these large angular-scale modes are more sensitive to the sky coverage and can therefore develop systematic fluctuations.
We therefore adopt $\ell_{\rm min}=10$ as the lower multipole cutoff, which entails only a minimal loss of statistical information given the large uncertainties at these scales.

\section{Selection Effects}
\label{app:selection}

\subsection{Galactic Latitude and Declination Cuts}
\label{app:GalDec}

\begin{figure*}
	\centering
	\includegraphics[width=0.32\textwidth]{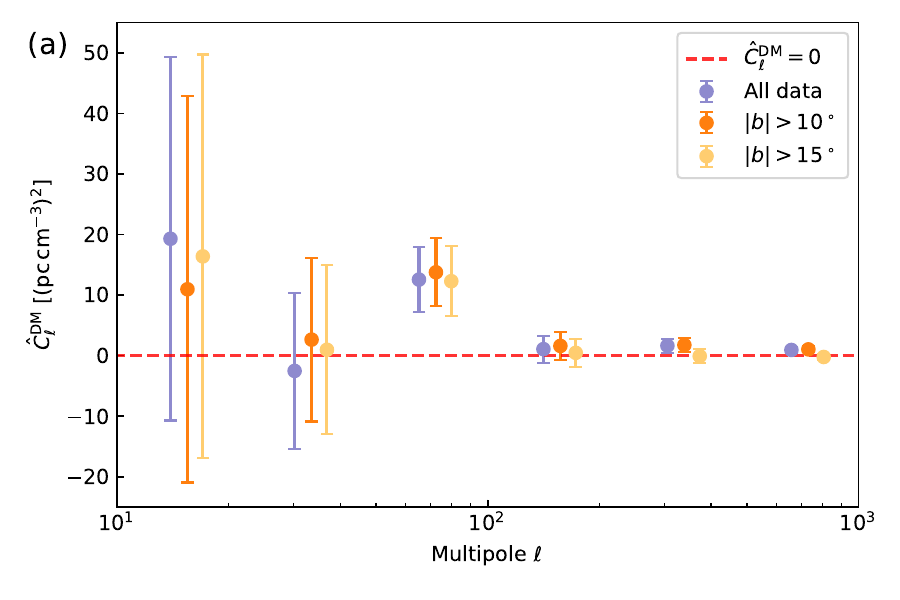}
    \includegraphics[width=0.32\textwidth]{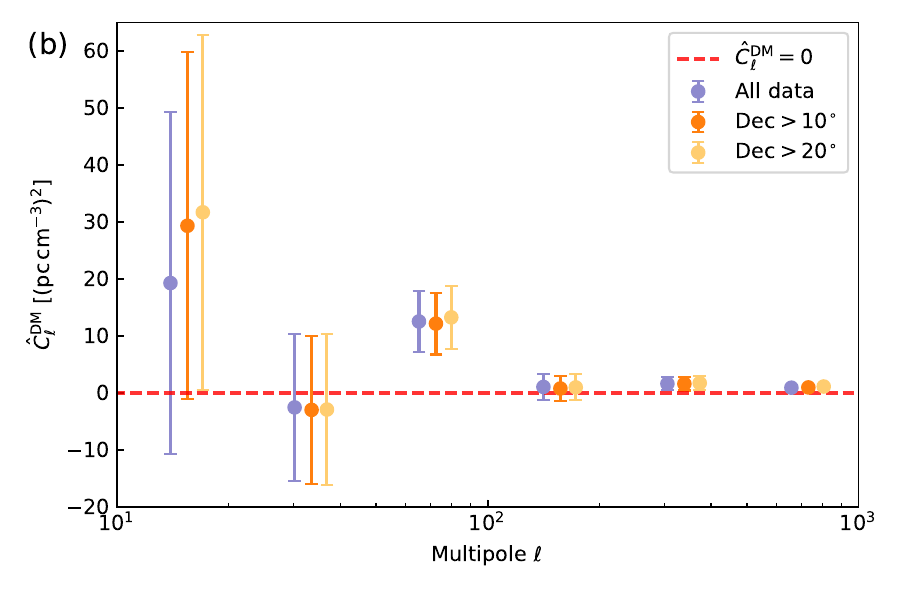}
    \includegraphics[width=0.32\textwidth]{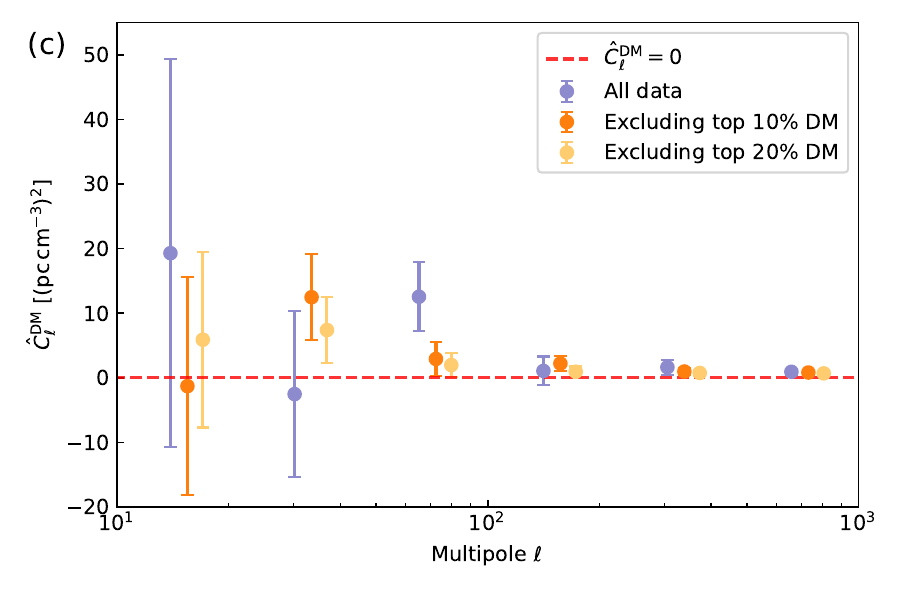}
	\caption{
        Comparison of the measured DM power spectra with different Galactic latitude cuts, declination cuts, and high-DM cuts.
        Panel (a): Galactic latitude cuts of $|b|>10^{\circ}$ and $|b|>15^{\circ}$.
        Panel (b): Declination cuts of ${\rm Dec}>10^\circ$ and ${\rm Dec}>20^\circ$.
        Panel (c): Exclusion of the highest 10\% and 20\% of observed DM values.
        The purple points represent the full sample, while the orange and yellow points correspond to the two cuts in each panel.
        The error bars represent the $1\sigma$ uncertainties, and the red dashed lines indicate zero power.
        Small horizontal offsets in $\ell$ are applied for visual clarity.
		}
	\label{fig:selection}
\end{figure*}

FRB detections can be affected by Galactic plasma-induced selection effects \citep{Patil2026ApJ} and by the declination-dependent sensitivity of the CHIME/FRB survey \citep{Chime2026ApJSa}.
To assess the impact of these selection effects, we perform the DM power spectrum measurements with different cuts on the Galactic latitude and declination. 
We consider FRB data satisfying $|b|>10^\circ$ and $|b|>15^\circ$ for the Galactic latitude cuts, as well as ${\rm Dec}>10^\circ$ and ${\rm Dec}>20^\circ$ for the declination cuts, and compare their results with the full sample.
As shown in Figure~\ref{fig:selection}(a) and (b), stricter Galactic latitude cuts suppress the measured bandpowers at nearly all angular scales, although the differences remain within the statistical uncertainties.
In contrast, the declination cuts have little impact on the measured power spectrum, with only modest changes at large angular scales.
We also examine the impact of the Galactic latitude cuts on cosmological parameter constraints and find only minor changes in the inferred parameters.

\subsection{High-DM Cuts}
\label{app:highDM}

To assess the sensitivity of the measured DM angular power spectrum to high-DM sources, we repeat the analysis after excluding the highest 10\% and 20\% of FRBs in the observed DM.
The resulting power spectra are compared with the full sample in Figure~\ref{fig:selection}(c).
Excluding high-DM sources leads to noticeable variations in several bandpowers, particularly at $\ell<100$.
These results indicate that high-DM sources contribute appreciably to the measured power spectrum, suggesting that the non-detection of high-DM FRBs due to observational selection effects could substantially affect the measured DM angular power spectrum.

\bibliographystyle{aasjournal}
\bibliography{ref}{}

\end{document}